\documentclass{ieeeaccess}
\usepackage{cite}
\usepackage{amsmath,amssymb,amsfonts}
\usepackage{bbm}
\usepackage{pseudocode}
\usepackage{color}

\usepackage{algorithmic}
\usepackage[colorlinks=true,
            linkcolor=red,
            urlcolor=blue,
            citecolor=blue]{hyperref}
            
\usepackage{graphicx}
\usepackage{textcomp}
\usepackage{lipsum}
\newcommand{\RNum}[1]{%
  \textup{\uppercase\expandafter{\romannumeral#1}}%
}

\DeclareRobustCommand*{\IEEEauthorrefmark}[1]{%
  \raisebox{0pt}[0pt][0pt]{\textsuperscript{\footnotesize #1}}%
}

\def\BibTeX{{\rm B\kern-.05em{\sc i\kern-.025em b}\kern-.08em
    T\kern-.1667em\lower.7ex\hbox{E}\kern-.125emX}}
\begin{document}
\history{Date of publication 10, 2018 , date of current version  10, 2018.}
\doi{10.1109/ACCESS.2018.2875677}

\title{Patient2Vec: A Personalized Interpretable Deep Representation of the Longitudinal Electronic Health Record}

\author{\uppercase{Jinghe Zhang}\IEEEauthorrefmark{1}, \IEEEmembership{Member, IEEE},
\uppercase{Kamran Kowsari}\IEEEauthorrefmark{1,4}\IEEEmembership{Member, IEEE}, 
\uppercase{James H. Harrison}\IEEEauthorrefmark{2,3,5}, 
\uppercase{Jennifer M. Lobo}\IEEEauthorrefmark{2,5}, 
and \uppercase{Laura E. Barnes}\IEEEauthorrefmark{1,4,5},
\IEEEmembership{Member, IEEE}\\}

\address[]{~\\}
\address{\IEEEauthorrefmark{1}Department of Systems and Information Engineering, 
University of Virginia, Charlottesville, VA 22904, USA}
\address{\IEEEauthorrefmark{2} Department of Public Health Sciences, University of Virginia, Charlottesville, VA 22904, USA}
\address{\IEEEauthorrefmark{4} Sensing Systems for Health Lab, University of Virginia, Charlottesville, VA 22904, USA}
\address{\IEEEauthorrefmark{3} Division of Laboratory Medicine Department of Pathology, University of Virginia, Charlottesville, VA 22904, USA}
\address{\IEEEauthorrefmark{5}Data Science Institute, University of Virginia, Charlottesville, VA 22904, USA}

\markboth
{Patient2Vec: A Personalized Interpretable Deep Representation of the Longitudinal Electronic Health Record}
{Patient2Vec: A Personalized Interpretable Deep Representation of the Longitudinal Electronic Health Record}

\corresp{Corresponding author: Laura E. Barnes (\href{mailto:lb3dp@virginia.edu}{lb3dp@virginia.edu})\\ 
This research was supported by a Jeffress Trust Award in Interdisciplinary Science. \\
Patient2Vec is shared as an open source tool at \url{https://github.com/BarnesLab/Patient2Vec}
}

\begin{abstract}
The wide implementation of electronic health record (EHR) systems facilitates the collection of large-scale health data from real clinical settings. Despite the significant increase in adoption of EHR systems, this data remains largely unexplored, but presents a rich data source for knowledge discovery from patient health histories in tasks such as understanding disease correlations and predicting health outcomes. However, the heterogeneity, sparsity, noise, and bias in this data present many complex challenges. This complexity makes it difficult to translate potentially relevant information into machine learning algorithms. 
In this paper, we propose a computational framework, \emph{Patient2Vec}, to learn an interpretable deep representation of longitudinal EHR data which is personalized for each patient. To evaluate this approach, we apply it to the prediction of future hospitalizations using real EHR data and compare its predictive performance with baseline methods. \emph{Patient2Vec} produces a vector space with meaningful structure and it achieves an AUC around~$0.799$ outperforming baseline methods. In the end, the learned feature importance can be visualized and interpreted at both the individual and population levels to bring clinical insights.
\end{abstract}

\begin{keywords}
Attention mechanism, gated recurrent unit, hospitalization, longitudinal electronic health record, personalization, representation learning.
% Enter key words or phrases in alphabetical 
% order, separated by commas. For a list of suggested keywords, send a blank 
% e-mail to keywords@ieee.org or visit \underline
% {http://www.ieee.org/organizations/pubs/ani\_prod/keywrd98.txt}
\end{keywords}

\titlepgskip=-15pt

\maketitle
% \fi
\IEEEpeerreviewmaketitle

\section{Introduction}
\label{sec1}
\PARstart{L}{}ongitudinal EHR data resemble text documents from many perspectives. A text document consists of a sequence of sentences, and a sentence is a sequence of words. Similarly, the longitudinal health record of a patient consists of a sequence of visits, and there is a list of clinical events, including diagnoses, medications, and procedures, that occur during a visit. Considering these similarities, representation learning methods for text documents in Natural Language Processing~(NLP) have great potential to be applied to longitudinal EHR data. 

Deep neural networks have become very popular in the NLP field and have been very successful in many applications, such as machine translation, question answering, text classification, document summarization, language modeling, etc.~\cite{yang2016hierarchical,kim2014convolutional,howard2018fine, lopez2017deep,cho2014learning,nobles2018identification,kowsari2017hdltex,Kowsari2018RMDL}.
These networks excel at complex language tasks because they are capable of identifying high-order relationships, the network structure can encode language structures, and they allow the learning of a hierarchical representation of the language, i.e., representations for tokens, phrases, and sentences, etc.

Among a variety of deep learning methods, Recurrent Neural Networks~(RNNs) have shown their effectiveness in NLP tasks because they have the ability to capture sequential information~\cite{Kowsari2018RMDL,kowsari2017hdltex,strobelt2018lstmvis,che2018recurrent} which is inherent in human language. Traditional neural networks assume that inputs are independent of each other, while an RNN computes the output based on the current input as well as the ``memory'' from the previous computation. Although vanilla RNNs are not good at capturing long-term dependencies, many variants have been proposed and validated that are effective in addressing this issue.

In the medical domain, it is critical that analytical results are interpretable, so that they can be understood and validated by a human with expert knowledge and so that knowledge captured by analysis can be used for process improvement. Traditional deep neural networks have the disadvantage that they lack interpretability. A substantial amount of work is ongoing to make sense of the ``black box'', and the attention mechanism~\cite{luong2015effective} is one of the more effective methods recently developed to make the output of these algorithms more interpretable. 

Health care is undergoing unprecedented change, and there is a great potential and demand for personalized care strategies. Personalized medicine, also called precision medicine, has previously focused on optimizing therapy to better fit the genetic makeup of the patient or the disease (e.g., the genetic susceptibility of cancer to specific chemotherapy strategies). The availability of EHR data and advances in machine learning create the potential for another type of personalization of healthcare. This type of personalization has become ubiquitous in our daily life. For example, customers have come to expect personalized search on Google and personalized product recommendations on Amazon and Netflix, based on their charactersitics and previous experiences with the systems. Personalization of healthcare processes, based on a patient's phenotype (physical and medical characteristics) and healthcare experiences as documented in the health record, may also improve "customer" satisfaction and it has the additional potential to improve healthcare efficiency, lower costs, and yield better outcomes. We believe that representation learning methods can capture a personalized representation of the important heterogeneities in patients' phenotypes and medical histories at the population-level, and make these representations available to drive healthcare decisions and strategies.

This research is based on RNN models and the attention mechanism with the objective of learning a personalized, interpretable, and complete representation of patients' medical records. Our proposed framework is capable of learning a personalized representation for each patient from a sequence of clinical events. A hierarchical attention mechanism learns personalized weights of clinical events, including hospital visits and the procedures that they contain. These weights allow us to interpret the relative importance and roles of clinical events in the learned representations both at individual and population levels. The ultimate goal is more accurate prediction and better insight into the critical elements of healthcare processes that can be used to improve healthcare delivery.

The rest of this paper is organized as follows: Section~\ref{sec2} summarizes the variants of RNNs and the attention mechanism, as well as their application to EHR data. Section~\ref{sec3} presents an overview of the proposed~\emph{Patient2Vec} representation learning framework, and Section~\ref{sec4} elaborates the details of the algorithms. In Section~\ref{sec5}, the proposed framework is evaluated for a prediction task and we compare its performance with other baseline methods. In addition to prediction performance, we further interpret the learned representations with visualizations on example patients and events. Finally, Section~\ref{sec5} provides a summary of this work.

\section{Related Work}
\label{sec2}
In this section, we present an overview of a gated recurrent unit, a type of RNN, which is capable of capturing long-term dependencies.
Then we briefly introduce attention mechanisms in neural networks that allow the network to attend to certain regions of data, which is inspired by the visual attention mechanism in humans.
Additionally, we summarize the RNN networks and attention mechanisms previously used to mine EHR data.

\subsection{Recurrent Neural Networks~(RNN)}
RNNs are expected to learn long-term dependencies by taking the previous state and the new input in the computation at the current time step $t$. However, vanilla RNNs are incapable of capturing the dependencies when the sequence is very long due to the vanishing gradient problem~\cite{lecun2015deep}. Many variants of the RNN network have been proposed to address this issue, and long short term memory (LSTM) is one of the most popular models used nowadays in NLP tasks~\cite{yogatama2017generative,young2017recent, kowsari2017hdltex,Kowsari2018RMDL,basaldella2018bidirectional, ghosh2016contextual}. 
\Figure[h](topskip=0pt, botskip=0pt, midskip=0pt)[width=.48\textwidth]{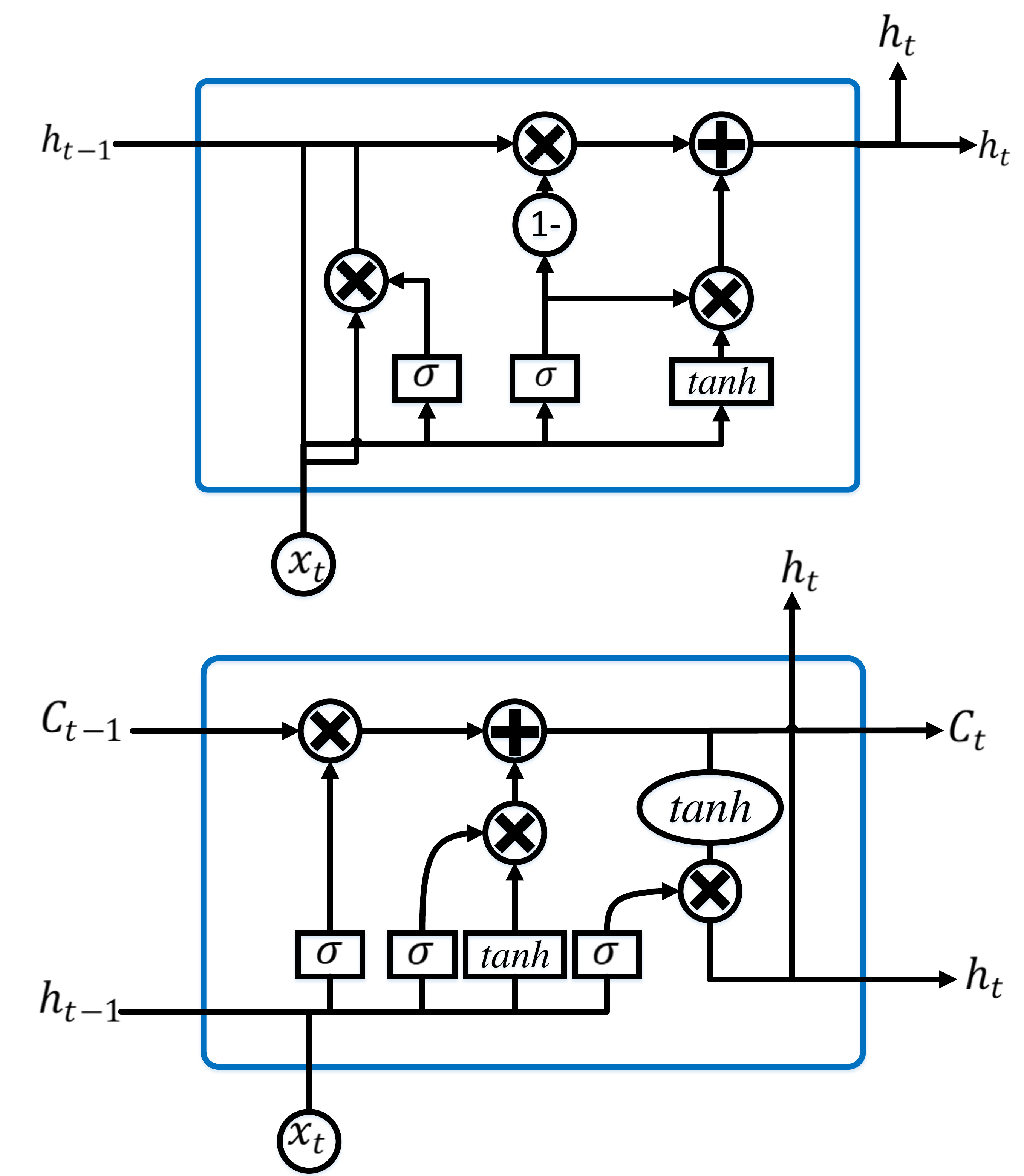}{The top figure is a GRU gating unit and bottom figure shows an LSTM unit~\cite{kowsari2017hdltex}\label{fig5_gru}}
\subsubsection{Gated Recurrent Unit (GRU)}
GRU is a simplified version of LSTM~\cite{kowsari2017hdltex}. The basic idea of GRU is to combat the vanishing gradient problem with a gating mechanism. Hence the general recurrent structure in GRU is identical to vanilla RNNs except that a GRU unit is used in the computation at each time step rather than a traditional simple recurrent unit. 

In general, a GRU cell has two gates, i.e., a reset gate~$r$ and an update gate~$z$. The reset gate is used to determine how to integrate the previous state into the computation of the current state, while the update gate determines how much the unit updates its activation. 

Given the input~$\mathbf{x}_t$ at time step~$t$, the reset gate~$r_t$ is computed as presented in Equation~\ref{eq5.9}

\begin{equation}
\label{eq5.9}
r_t = \sigma(\mathbf{U}_r \mathbf{x}_t + \mathbf{W}_r \mathbf{s}_{t-1})
\end{equation}
where~$\mathbf{U}_r$ and~$\mathbf{W}_r$ are the weight matrices of the reset gate and~$\mathbf{s}_{t-1}$ is the hidden activation at time step~$t-1$. A similar computation is performed for the update gate~$z_t$ at time step~$t$, shown in Equation~\ref{eq5.10}

\begin{equation}
\label{eq5.10}
z_t = \sigma(\mathbf{U}_z \mathbf{x}_t + \mathbf{W}_z \mathbf{s}_{t-1})
\end{equation}
where~$\mathbf{U}_z$ and~$\mathbf{W}_z$ are the weight matrices of update gate. The current hidden activation~$\mathbf{h}_t$ is computed by

\begin{equation}
\label{eq5.11}
\mathbf{h}_t = (1-z_t) \mathbf{h}_{t-1} + z_t \tilde{\mathbf{h}}_t
\end{equation}
where $\tilde{\mathbf{h}}_t$ is the candidate activation at time step $t$. The computation of $\tilde{\mathbf{h}}_t$ is presented in Equation~\ref{eq5.12}

\begin{equation}
\label{eq5.12}
\tilde{\mathbf{h}}_t = \tanh(\mathbf{W} \mathbf{x}_t + \mathbf{U} (r_t \odot \mathbf{h}_{t-1}))
\end{equation}
where $\mathbf{U}$ and $\mathbf{W}$ are weight matrices and $\odot$ represents element-wise multiplication. Figure~\ref{fig5_gru} presents a graphical illustration of the GRU~\cite{kowsari2017hdltex} and one unit of LSTM.

GRU is capable of learning long-term dependencies~\cite{yue2018residual} due to the additive component of update from $t$ to $t+1$ in the gating mechanism. Consequently, important features will be carried forward in the input stream while irrelevant information will be dropped. When the reset gate is $0$, the network is forced to drop previous states and reset with current information. Moreover, the method provides shortcuts such that the error is easily backpropagated without vanishing too quickly~\cite{cho2014learning, chung2014empirical}. Hence, the GRU is well-suited to learn long-term dependencies in sequence data.

\subsubsection{Long Short-Term Memory (LSTM)}
An LSTM unit is similar to a GRU, but with one more gate in an LSTM unit~(as shown in Figure~\ref{fig5_gru}). LSTM also preserves long term dependencies more effectively than basic RNN. This is particularly useful to overcome the vanishing gradient problem~\cite{pascanu2013difficulty}. Although LSTM has a chain-like structure similar to RNN,
LSTM uses multiple gates to carefully regulate the amount of information that will be allowed into each node state. Figure~\ref{fig5_gru} shows the basic cell of an LSTM model. A step by step explanation of an LSTM cell is as following:\\
Input gate:
\begin{equation}
    i_{t}=\sigma(\mathbf{W}_{i}[\mathbf{x}_{t},\mathbf{h}_{t-1}]+b_{i}), \label{eq:lstm1}
\end{equation}
Candid memory cell value:
\begin{equation}
    \tilde{\mathbf{C}_{t}}=\tanh(\mathbf{W}_{c}[\mathbf{x}_{t},\mathbf{h}_{t-1}]+b_{c}), \label{eq:lstm2} 
\end{equation}
Forget gate activation:
\begin{equation}
    f_{t}=\sigma(\mathbf{W}_{f}[\mathbf{x}_{t},\mathbf{h}_{t-1}]+b_{f}), \label{eq:lstm3}
\end{equation}
New memory cell value:
\begin{equation}
    \mathbf{C}_{t}= i_{t}* \tilde{\mathbf{C}_{t}}+f_{t} \mathbf{C}_{t-1}, \label{eq:lstm4}
\end{equation}
Output gate value:
\begin{equation}
    o_{t}= \sigma(\mathbf{W}_{o}[\mathbf{x}_{t},\mathbf{h}_{t-1}]+b_{o}), \label{eq:lstm5}
    \end{equation}
\begin{equation}
    \mathbf{h}_{t}=o_{t}\tanh(\mathbf{C}_{t}),\label{eq:lstm6}
\end{equation}

In the above description all~$\mathbf{b}$ represent bias vectors, all~$\mathbf{W}$ represent weight matrices, and~$\mathbf{x}_{t}$ is used as input to the memory cell at time~$t$. Also,the~$i,c,f,o$ indices refer to input, cell memory, forget and output gates respectively. An RNN can be biased when later words are more influential than the earlier ones.

Empirically, LSTM and GRU achieve comparable performance in many tasks but there are fewer parameters in a~GRU, which makes it a little faster to learn and able to generalize with fewer data~\cite{wildml}. 

\subsection{Attention Mechanism}
Attention mechanisms, inspired by the visual attention system found in humans, have become popular in deep learning. Attention allows the network to focus on certain regions of data, while perceiving other regions with ``low resolution''. In addition to higher accuracy, it also facilitates the interpretation of learned representations. We elaborate an attention mechanism on an RNN network, and Figure~\ref{fig5_attention} presents a graphical illustration.

\Figure[t!](topskip=0pt, botskip=0pt, midskip=0pt)[width=.48\textwidth]{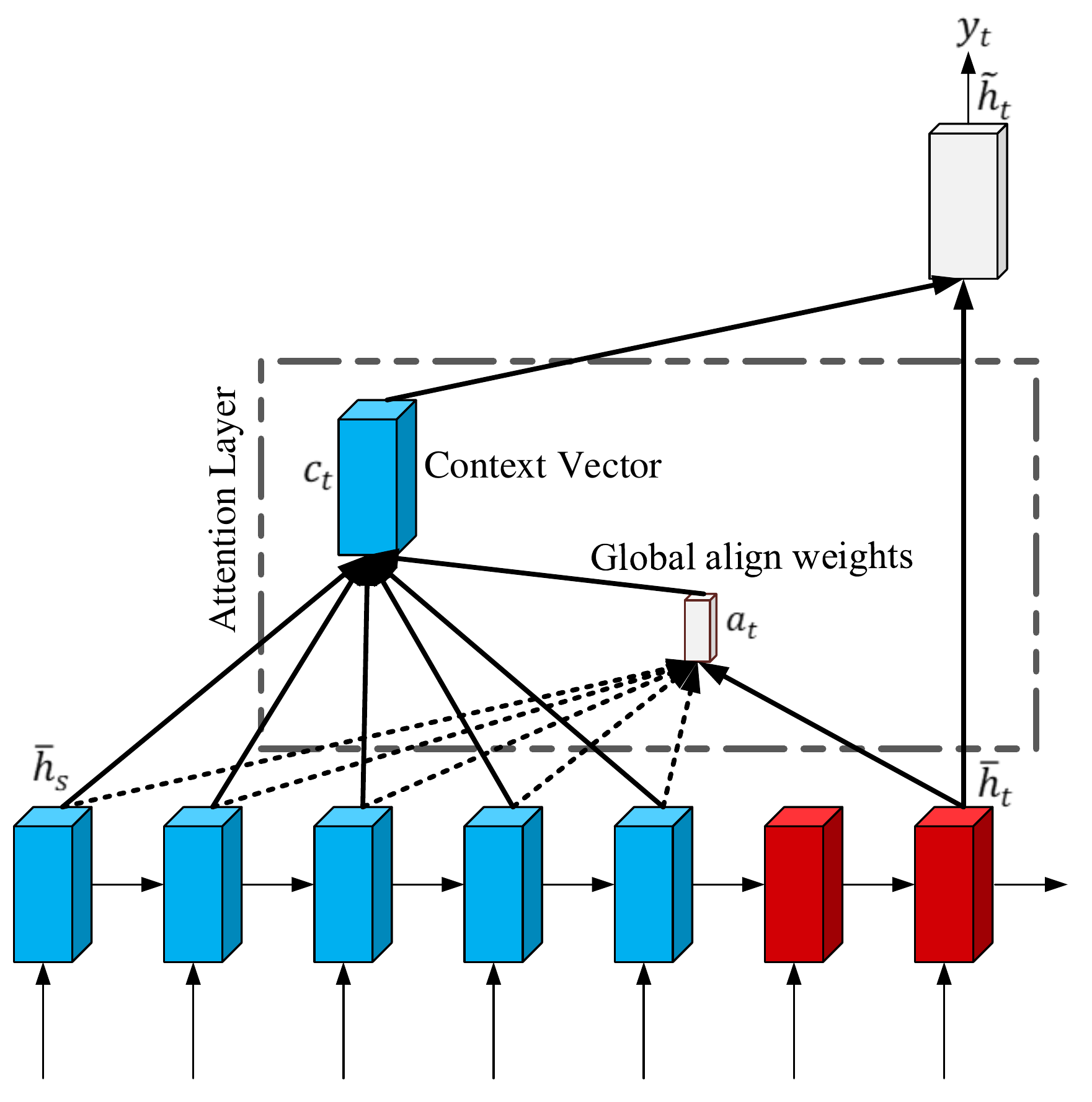}{The global attention model\label{fig5_attention}}

According to Figure~\ref{fig5_attention}, a variable-length weight vector $\mathbf{\alpha}$ is learned based on hidden states~\cite{luong2015effective}. Then a global context vector is computed based on weights $\mathbf{\alpha}$ and all the hidden states to create the final output. Equation~\ref{eq5.13} presents the computation of the weight vector $\mathbf{\alpha} = \{\alpha_1, \alpha_2, \cdots, \alpha_T\}$, where $T$ is the length of the sequence

\begin{equation}
\label{eq5.13}
\alpha_1, \alpha_2, \cdots, \alpha_T = f(\mathbf{W}_{\alpha} \mathbf{h} + b_{\alpha})
\end{equation}
and where $f$ is a nonlinear activation function, usually $softmax$ or $\tanh$. Then, the context vector $c$ is constructed as:

\begin{equation}
\label{eq5.14}
\mathbf{c} = \displaystyle\sum_{t=1}^T \alpha_t \mathbf{h}_t
\end{equation}

Thus, the network puts more attention on the important features for the final prediction which can improve the model performance. An additional benefit is that the weights can be utilized to understand the importance of features such that the models are more interpretable. The attention mechanism has been introduced to both Convolutional Neural Networks~(CNNs) and RNNs for various tasks and has achieved many successes in the fields of computer vision and NLP~\cite{luong2015effective, mnih2014recurrent, rush2015neural}. 

\subsection{Deep Learning in EHR Data}
Previous studies on EHR data mainly use statistical methods or traditional machine learning techniques. Recently researchers have started adapting deep learning approaches to this data~\cite{ma2018health,rajkomar2018scalable}, including textual notes, temporal measurements of laboratory testing in the Intensive Care Unit~(ICU), and longitudinal data in patient populations. Here, we summarize deep learning research in mining EHR data and focus on the studies using RNN-based models. 

Hospitalized patients, especially patients in ICUs, are continuously monitored for cardiac, respiratory, and other physical functions, creating a large volume of sequential data in multiple dimensions. These measurements are utilized by physicians to make diagnostic and treatment decisions. The functions monitored may change over time and monitoring may be irregular, based on a patient's condition. It is very challenging for traditional machine learning methods to mine this multivariate time series data considering missing values, varying length, and irregular, non-simultaneous sampling.~\textit{Lipton et al.}~\cite{lipton2015learning} trained an LSTM with a replicated target to learn from these sequence data and used this model to make predictions of diagnoses. The data used in this research are time series of clinical measurements with continuous values, and the LSTM models outperformed logistic regression and MLP. ~~\textit{Che et al.}~\cite{che2016recurrent} developed a GRU-based model to address missing values in multivariate time series data, in which the missing patterns are incorporated for improved prediction performance. This work has been applied to the Medical Information Mart for Intensive Care III~(MIMIC-III) clinical database to demonstrate its effectiveness in mining time series of clinical measurements with missing values~\cite{johnson2016mimic}. Longitudinal EHR data including clinical events, such as diagnoses, medications, and procedures is also a potentially rich resource for predictive modeling.~\textit{Choi~et al.}~\cite{choi2016doctor} analyze this data with a GRU network to forecast future clinical events, and it achieves a better prediction performance than comparison models such as logistic regression and MLP.

Difficulty in interpreting model behavior is one of the major drawbacks of using deep learning to mine EHR data. Some attempts have been made to address this issue.~\textit{Che~et al.}~\cite{che2016interpretable} propose an interpretable mimic learning method which trains a mimic gradient boosting trees model to utilize predicted labels or features learned by deep learning models for final prediction~\cite{friedman2001greedy}. Then the feature importances learned by the tree-based models are used for knowledge discovery. Attention mechanisms have been introduced recently to improve the interpretability of the prediction results of deep learning models in health analytics. ~\textit{Choi~et al.}~\cite{choi2016retain} develop an interpretable model with two levels of attention weights learned from two reverse-time GRU models, respectively. The experimental results on EHR data indicate comparable prediction performance with conventional GRU models but more interpretable results. Our work continues the attempt to use attention mechanisms to improve the interpretability of RNN-based models. 

\section{\emph{Patient2Vec} System Model}
\label{sec3}
In this section, we provide an overview of the proposed hierarchical representation learning framework. This framework uses deep recurrent neural networks to capture the complex relationships between clinical events in the patient's EHR data and employs the attention mechanism to learn a personalized representation and to obtain relative feature importance. The proposed representation learning framework contains four steps and is presented graphically in Figure~\ref{fig5_system}.

\Figure[t!](topskip=0pt, botskip=0pt, midskip=0pt)[width=.48\textwidth]{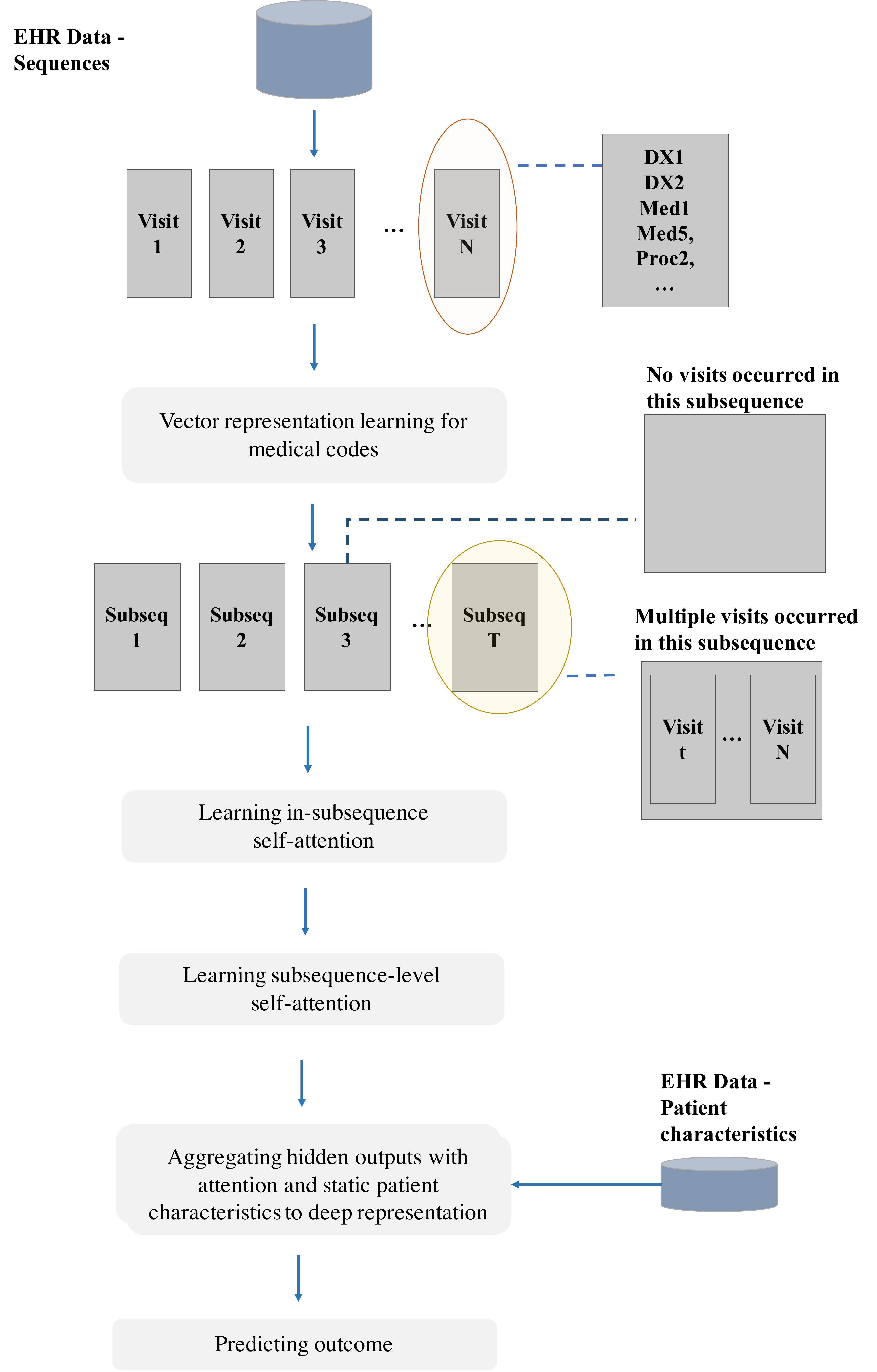}{The \emph{Patient2Vec} representation learning framework \label{fig5_system}}

\subsection{Learning vector representations of medical codes} EHR data consists primarily of records of outpatient and inpatient visits to healthcare providers. These visit records include multiple clinical codes for diagnoses, symptoms, procedures, therapies, and other observations and events that occurred during the visit. Here, we treat the set of medical codes associated with a visit as a sentence consisting of words, except that there is no ordering in the words. Thus, we adopt the word2vec approach to construct a vector to represent each medical code. 

\subsection{Learning within-subsequence self-attention} Clinical visits are represented as the set of vectors for the codes associated with the visit. Because closely-spaced visits are usually related clinically, we employ a time window to split the sequence of visits into multiple subsequences of equal length. A subsequence might contain multiple visits if they occurred within the same time window, or there might be no visits during a particular time window yielding an empty subsequence. Thus we transform the original sequence of irregularly-spaced visits into a sequence of subsequences with equal intervals, which is preferable for recurrent neural networks. The width of the subsequence window defines the time granularity of the method and its optimal width is related to the acuity (i.e., stability) of the clinical characteristics involved in the predication task. In future work it may be possible to define the relationship between clinical acuity and optimal subsequence width, or develop methods for learning an optimal width for a defined prediction task.

Because all medical events occurring within a subsequence are unlikely to contribute equally to the prediction of the target outcome, we cannot aggregate them with equal weights. Instead, we employ a self-attention mechanism which trains the network to learn the weights. 

\subsection{Learning subsequence-level self-attention}  Given a sequence of subsequences with embedded medical codes, we are able to input it into a recurrent neural network to capture the temporal dependencies between events. However, the subsequences of visits are not contributing equally to the outcome. Hence, we employ another level of attention to learn the weights of the subsequences by the network itself for the outcome prediction. 

\subsection{Constructing aggregated deep representation}  Given the learned weights and hidden outputs, we aggregate them into one universal vector for a comprehensive representation. In this step, the static information, such as age, gender, previous hospitalization history is added as extra features, to get a complete representation of a patient.

\subsection{Predicting outcome} Given the complete vector representation of a patient's EHR data, we add a logistic regression layer at the end for the prediction of outcome.

\section{Patient2Vec Representation Learning Algorithm}
\label{sec4}
\begin{figure*}
\centering
\includegraphics[width=0.75\textwidth]{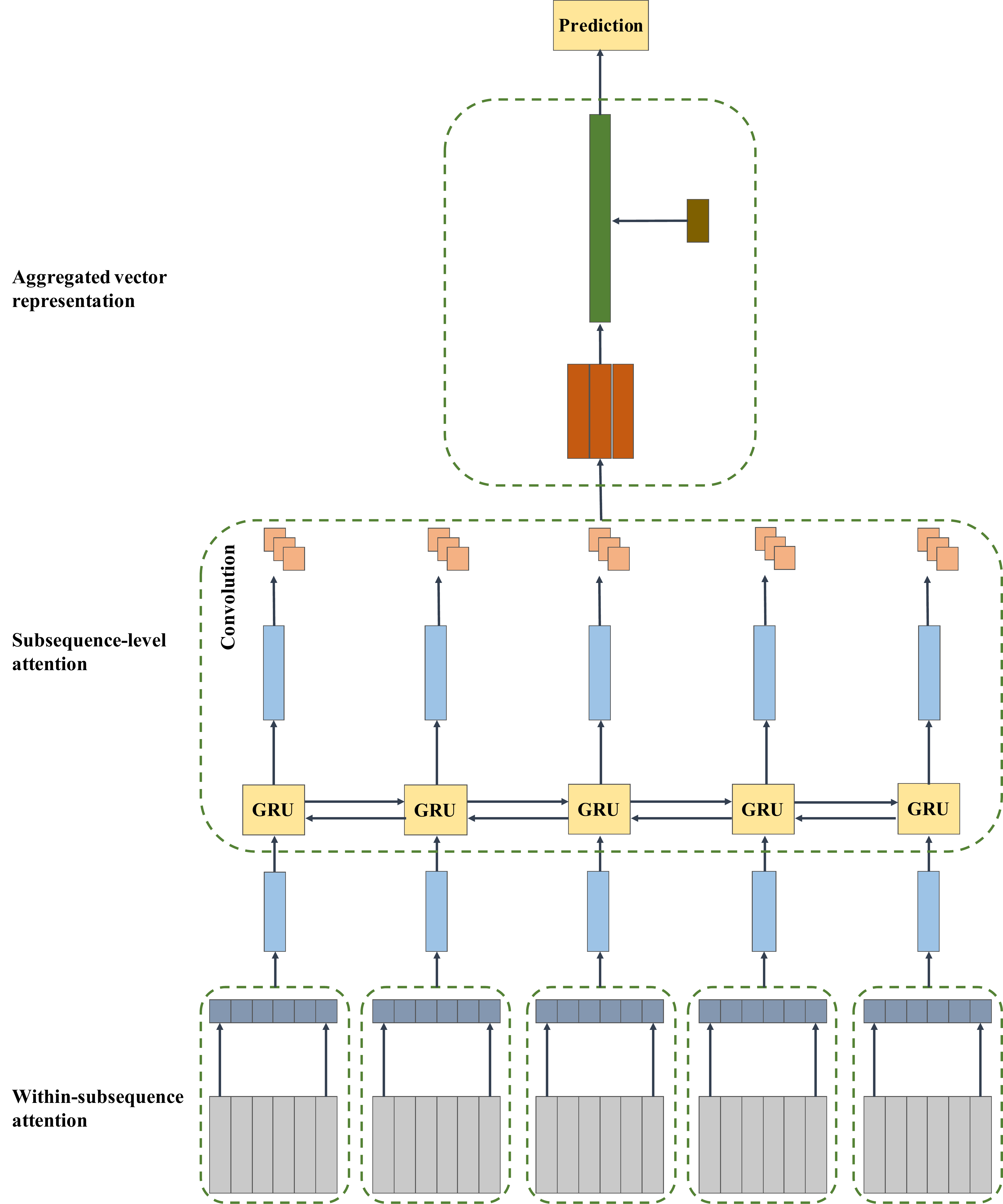}
\caption{A graphical illustration of the network in the \emph{Patient2Vec} representation learning framework}
\label{fig5_network}
\end{figure*}

In this section, we present the details of the proposed representation learning framework, which is based on a GRU network and a hierarchical attention mechanism. Figure~\ref{fig5_network} presents the structure of the proposed network with attention.

The proposed framework consists of five parts presented in the following:~\RNum{1}) Learning vector representations of medical codes,~\RNum{2}) Learning within-subsequence self-attention,~\RNum{3}) Learning subsequence-level self-attention,~\RNum{4}) Constructing aggregated deep representation,~\RNum{5}) Predicting outcome.

\subsection{Learning vector representations of medical codes}
Given a patient's raw EHR data, a sequence of visits, we observe that a visit usually contains multiple medical codes. Hence, it is feasible to learn a vector to represent the medical code by capturing the relationships between the codes. In this work, we employ the classical word2vec algorithm, skip-gram. The basic idea of skip-gram is to learn a vector to represent each word such that the probability of the context to predict based on the target word is maximized. Hence, the vectors of similar words are close to each other in the learned feature space. In the skip-gram model, the vectors are learned by training a shallow neural network to predict the context words given an input word. Similarly, in our problem, the input is a medical code and the target to predict are the medical codes occurred in the same visit. 

Hence, each subsequence is a matrix consisting of the vectors of medical codes occurred during this associated time window. 

\subsection{Learning within-subsequence self-attention}
Given a sequence of subsequences encoded by vectors of medical codes, this step employs the within-subsequence attention which allows the network itself to learn the weights of vectors in the subsequence according to its contribution to the prediction target. 

Here, we denote the sequence of patient~$i$ as~$\mathbf{s}^{(i)}$, and~$\mathbf{v}_t^{(i)}$ denotes the $t$th subsequence in sequence $\mathbf{s}^{(i)}$, where~$t \in \{1, 2, \cdots, T\}$. Thus, $\mathbf{s}^{(i)}=\{\mathbf{v}_1^{(i)}, \cdots, \mathbf{v}_t^{(i)}, \cdots, \mathbf{v}_T^{(i)}\}$. To simplify the notation, we omit $i$ in the following explanation. Subsequence $\mathbf{v}_t \in \mathbb{R}^{n\times d}$ is a matrix of medical codes such that $\mathbf{v}_t = \{v_{t_{1}}, v_{t_{2}}, \cdots, v_{t_{j}}, \cdots, v_{t_{n}}\}$, where $v_{t_{j}} \in \mathbb{R}^{d}$ is the vector representation of the $j$th medical code in the $t$th subsequence $\mathbf{v}_t$ and there are $n$ medical codes in a subsequence. In real EHR data, it is very likely that the numbers of medical codes in each visit or time window are different, thus, we utilize the padding approach to obtain a consistent matrix dimensionality in the network.

To assign attention weights, we utilize the one-side convolution operation with a filter $\mathbf{\omega}^\alpha \in \mathbb{R}^{d}$ and a nonlinear activation function. Thus, the weight vector $\mathbf{\alpha}_t$ is generated for medical codes in the subsequence $\mathbf{v}_t$, presented in Equation~\ref{eq5.1}.

\begin{equation}
\label{eq5.1}
\mathbf{\alpha}_{t} = \tanh(Conv(\mathbf{\omega}^\alpha, \mathbf{v}_t))
\end{equation}
where $\mathbf{\alpha}_{t} = \{\alpha_{t_{1}}, \alpha_{t_{2}}, \cdots, \alpha_{t_{n}}\}$, and $\mathbf{\omega}^\alpha \in \mathbb{R}^d$ is the weight vector of the filter. The convolution operation $Conv$ is presented in Equation~\ref{eq5.2}.

\begin{equation}
\label{eq5.2}
\tilde{\alpha}_{t_{j}} = (\mathbf{\omega}^\alpha) ^\intercal \mathbf{v}_{t_{j}} + b^\alpha
\end{equation}
where $b^\alpha$ is a bias term. Then, given the original matrix $\mathbf{v}_t$ and the learned weights $\mathbf{\alpha}_t$, an aggregated vector $\mathbf{x}_t \in \mathbb{R}^{d}$ is constructed to represent the $t$th subsequence, presented in~\ref{eq5.3}. 

\begin{equation}
\label{eq5.3}
\mathbf{x}_t = \displaystyle\sum_{j=1}^n\alpha_{t_{j}} v_{t_{j}} 
\end{equation}
Given Equation~\ref{eq5.3}, we obtain a sequence of vectors,~$\mathbf{x}=\{\mathbf{x}_1,\mathbf{x}_2,\cdots,\mathbf{x}_t,\cdots,\mathbf{x}_T\}$, to represent a patient's medical history. 

\subsection{Learning subsequence-level self-attention}
Given a sequence of embedded subsequences, this step employs the subsequence-level attention which allows the network itself to learn the weights of subsequences according to their contribution to the prediction target. 

To capture the longitudinal dependencies, we utilize a bidirectional GRU-based RNN, presented in Equations~\ref{eq5.4}.

\begin{equation}
\label{eq5.4}
\mathbf{h}_1, \cdots, \mathbf{h}_t, \cdots, \mathbf{h}_T = GRU(\mathbf{x}_1, \cdots, \mathbf{x}_t, \cdots, \mathbf{x}_T)
\end{equation}
where $\mathbf{h}_t \in \mathbb{R}^k$ represents the output by the GRU unit at the $t$th subsequence. Then, we introduce a set of linear and softmax layers to generate $M$ hops of weights $\mathbf{\beta} \in \mathbb{R}^{M\times T}$ for subsequences. Then, for the hop $m$

\begin{align}
\label{eq5.5}
\gamma_{mt} &= (\mathbf{w}_m^{\beta})^ \intercal \mathbf{h}_t + b^{\beta}\\
\beta_{m_1}, \cdots, \beta_{m_T} &=\label{eq5.5.1}\\ soft&max(\gamma_{m_1}, \cdots, \gamma_{m_t}, \cdots, \gamma_{m_T})\nonumber
\end{align}
where $\mathbf{w}_m^{\beta} \in \mathbb{R}^{k}$. Thus, with the subsequence-level weights and hidden outputs, we construct a vector $\mathbf{c}_m \in \mathbb{R}^k$ to represent a patient's medical visit history with one hop of subsequence weights, presented in the following Equation~\ref{eq5.6}.

\begin{equation}
\label{eq5.6}
\mathbf{c}_m = \displaystyle\sum_{t=1}^{T} \beta_{mt} \mathbf{h}_t
\end{equation}

Then, a context vector $\mathbf{c} \in \mathbb{R}^{M \times k}$ is constructed by concatenating $\mathbf{c}_1$, $\mathbf{c}_2$, $\cdots$, $\mathbf{c}_M$.

\subsection{Constructing aggregated deep representation}
Given the context vector $\mathbf{c}$, this step integrates the patients characteristics $\mathbf{a} \in \mathbb{R}^q$ into the context vector for a complete vector representation of the patient's EHR data. In this research, the patient characteristics include demographic information and some static medical conditions, such as age, gender, and previous hospitalization. Thus, an aggregated vector is constructed, $\mathbf{c}' \in \mathbb{R}^{M \times k+q}$, by adding $\mathbf{a}$ as additional dimensions to the context vector $\mathbf{c}$.

\subsection{Predicting outcome}
Given the vector representation of the complete medical history and characteristics of patients, $\mathbf{c}'$, we add a linear and a softmax layer for the final outcome prediction, as presented in Equation~\ref{eq5.7}.

\begin{equation}
\label{eq5.7}
\hat{y} = softmax({\mathbf{w}^{c}}^\intercal \mathbf{c}' + b^{c})
\end{equation}

To train the network, we use cross-entropy as the loss function, presented in Equation~\ref{eq5.8}.
\begin{equation}
    \begin{split}
        L =& -\frac{1}{N} \displaystyle\sum_{n=1}^N y_i log(\hat{y}_i) + (1-y_i) log(1-\hat{y}_i)\\ &+\frac{1}{N}\displaystyle\sum_{n=1}^N|| \mathbf{\beta} \mathbf{\beta} ^ \intercal -\mathbf{I}||_F^2\label{eq5.8}
    \end{split}
\end{equation}

\noindent where $N$ is the total number of observations. Here, $y_i$ is a binary variable in classification problems, while model output $\hat{y}_i$ is real-valued. The second term in Equation~\ref{eq5.8} is to penalize redundancy if the attention mechanism provides similar subsequence weights for different hops of attention, which is derived from~\cite{lin2017structured}. This penalty term encourages the multiple hops to focus on diverse areas and each hop focuses on a small area.

Thus, we obtain a final output for the prediction of outcomes and a complete personalized vector representation of the patient's longitudinal EHR data. 

\section{Evaluation}
\label{sec5}
\subsection{Background}
\label{sec5.5.1}
Although health care spending has been a relatively stable share of the Gross Domestic Product (GDP) in the United States since~$2009$, the costs of hospitalization, the largest single component of health care expenditures, increased by~$4.1\%$ in~$2014$~\cite{cost_hcup}. Unplanned hospitalization is also distressing and can increase the risk of related adverse events, such as hospital-acquired infections and falls~\cite{wallace2014risk,de2008incidence}. Approximately $40\%$ hospitalizations in the United Kingdom are unplanned and are potentially avoidable~\cite{purdey2012predicting}. One important form of unplanned hospitalization is hospital re-admissions within 30 days of discharge, which is financially penalized in the United States. Early interventions targeted to patients at risk of hospitalization could help avoid unplanned admissions, reduce inpatient health care cost and financial penalties for providers, and reduce emergency department congestion~\cite{canada_admission}.

In this research, we apply our proposed representation learning framework to the risk prediction of future hospitalization. Many studies have been conducted by researchers to predict the risk of $30$-day readmission, or the admission risk of a particular population, such as patients with Ambulatory Care Sensitive Conditions (ACSCs), patients with heart failure, etc.~\cite{zheng_predictive_2015, kansagara2011risk, giamouzis2011hospitalization, prescott1997gender}. Here, we focus on the general population and the objective is to predict the risk of all-cause hospitalization using longitudinal EHR data.

\subsection{Experimental Design}
\label{sec5.5.2}
In this research, we use de-identified EHR data from the University of Virginia Health System covering $75$ months beginning in September $2010$. This dataset contains $2{,}343{,}651$ inpatient and outpatient visits of $473{,}915$ distinct patients. We extracted visit data with diagnosis, medication, and procedure codes. 

\begin{figure}
\centering
\includegraphics[width=0.48\textwidth]{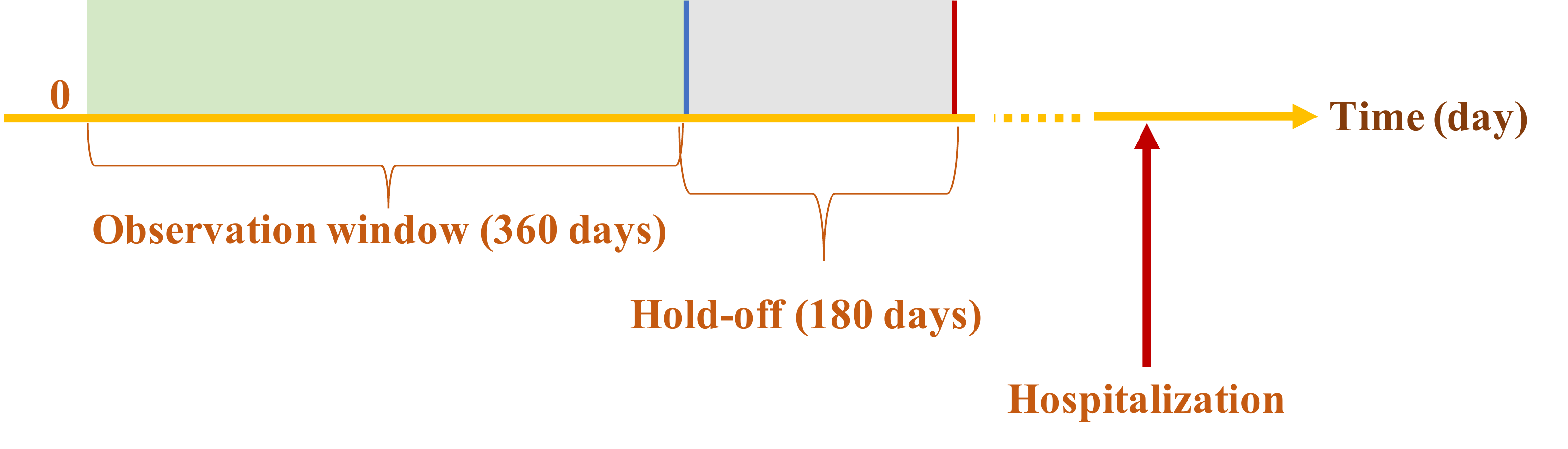}
\caption{A graphical illustration of the experimental setting for the risk prediction of hospitalization}
\label{fig5.setting}
\end{figure}

We defined the observation window and prediction period to validate the proposed method. We first extract all patients with a medical record of at least~$1.5$ years, where the first year is the observation window and the medical records in this time window are used for feature construction. The following~$6$ months is the hold-off period for the purpose of early detection. For the positive class, we take all patients who have hospitalization after the first $1.5$ years in their medical history, while the negative class consists of patients who have no hospitalization after $1.5$ years. To better illustrate the experimental setting, we present the observation window, hold-off and onset of outcome event in Figure~\ref{fig5.setting}. 
Here, the medical codes include diagnosis, medication, and procedure codes, and a vector representation is learned for each code. In this dataset, diagnoses are primarily coded in ICD-$9$ and a small portion is ICD-$10$ codes, while procedures are mainly using CPT codes with a few ICD-$9$ procedure codes. The codes of medications are using the pharmaceutical categories. Overall, there are~$94$ distinct medication categories,~$34{,}419$ distinct diagnoses codes, and~$7{,}895$ distinct procedure codes in the EHR data. The dimension of the learned vectors of medical codes is set to~$100$. Medical codes that appear in less than~$50$ patients medical records are excluded as rare events.

\begin{figure}
\centering
\includegraphics[width=0.49\textwidth]{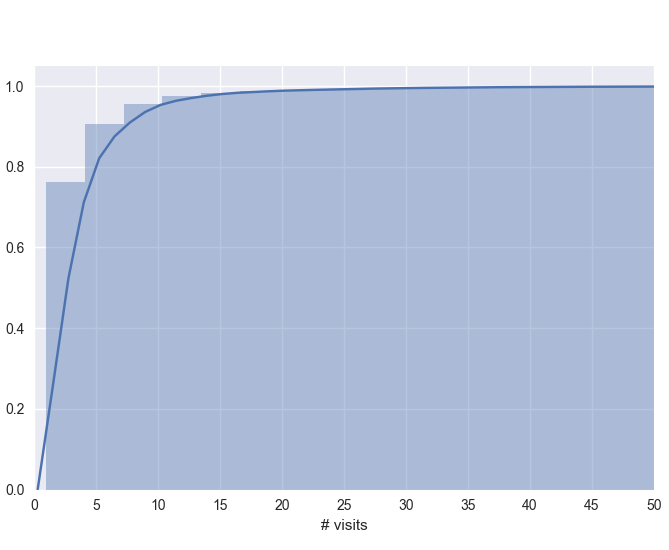}
\caption{The cumulative histogram and density plot of patients' numbers of visits}
\label{fig5_numvisit}
\end{figure}

To construct the subsequences of medical codes, we use $l$ days as the time window. Figure~\ref{fig5_numvisit} presents the cumulative histogram and density plot of the numbers of visits in the observation window, and we observe that the majority of patients have a small number of visits during the observation window (less than~$25\%$ of patients have more than~$4$ visits). Thus, we set~$l$ to~$90$ days, which split the observation window into~$4$ subsequences. 

Within each subsequence, the number of distinct medical codes were computed and patients with more medical codes in a subsequence than the $95\%$ quantile were excluded from the dataset. Overall, there are $8{,}841$ and $89{,}101$ patients in the target and control groups, respectively. Each group is randomly split into training, validation and testing sets with a 7:1:2 ratio. Thus,~$70\%$ are used for training, another~$20\%$ is used for testing, and the rest~$10\%$ are used for parameter tuning and early stopping. The stochastic gradient descent algorithm is used in training to minimize the cross-entropy loss function, shown in Equation~\ref{eq5.8}.

To evaluate the proposed representation learning framework, we compare the prediction performance of the proposed model with baseline approaches as follows.

\subsubsection{Logistic regression (LR)} The inputs are the aggregated counts of grouped medical codes over the entire observation window. Since the dimensionality of raw medical codes is huge, AHRQ clinical classifications of diagnoses and procedures are used to achieve a more general clustering of medical codes~\cite{ahrq}. The medication codes are the pharmaceutical classes. Furthermore, patient characteristics and previous inpatient visit are also considered, where age and gender are demographic information, and a binary indicator is utilized to represent the presence of the previous hospitalization. Hence, the input is a~$436$-dimensional vector representing a patient's medical history and characteristics. 
\begin{table*}[h]
\centering
\caption{The predictive performance of baselines and the proposed \emph{Patient2Vec} framework}
\label{tab5_result}
\begin{tabular}{lcccc}
\hline
Methods~~~~~~~~~~~~& Sensitivity & Specificity & AUC & F2 score \vspace{2pt}\\ \cline{2-5} 
LR & $0.637 \pm 0.010$ & $0.728 \pm 0.003$& $0.721 \pm 0.006$ & $0.434 \pm 0.006$\vspace{2pt}\\
MLP & $0.727 \pm 0.013$ & $0.617 \pm 0.004$ & $0.713 \pm 0.007$ & $0.423 \pm 0.007$\vspace{2pt}\\
RETAIN & $0.553 \pm 0.012$ & $0.710 \pm 0.003$ & $0.663 \pm 0.007$ & $0.370 \pm 0.008$\vspace{2pt}\\
FRNN-MGE & $0.636 \pm 0.012$ & $0.739 \pm 0.004$ & $0.759\pm 0.006$ & $0.438 \pm 0.009$ \vspace{2pt}\\
BiRNN-MGE & $0.600 \pm 0.012$ & $\mathbf{0.777 \pm 0.003}$ & $0.768 \pm 0.007$ & $0.439 \pm 0.009$\vspace{2pt}\\
FRNN-MVE & $0.753 \pm 0.011$ & $0.676 \pm 0.004$ & $0.785 \pm 0.006$ & $0.470 \pm 0.008$\vspace{2pt}\\ 
BiRNN-MVE & $0.724 \pm 0.010$ & $0.707 \pm 0.003$ & $0.788 \pm 0.005$ & $0.473 \pm 0.008$ \vspace{2pt}\\
\textbf{Patient2Vec} & $\mathbf{0.769 \pm 0.010}$ & $0.694 \pm 0.004$ & $\mathbf{0.799 \pm 0.005}$ & $\mathbf{0.492 \pm 0.007}$ \vspace{2pt}\\ \hline
\end{tabular}
\end{table*}

\subsubsection{Multi-layer perceptron (MLP)} A multi-layer perceptron is trained to predict hospitalization using the same inputs for logistic regression. Here, we use a one hidden layer MLP with~$256$ hidden nodes. 

\subsubsection{Forward RNN with medical group embedding (FRNN-MGE)} We split the sequence into subsequences with equal interval $l$. The input at each step is the counts of medical groups within the associated time interval, and the patient characteristics are appended as additional features in the final logistic regression step. Here, the RNN is a forward GRU~(or LSTM~\cite{chung2014empirical}) with one hidden layer and the size of the hidden layer is~$256$.
\subsubsection{Bidirectional RNN with medical group embedding (BiRNN-MGE)} The inputs used for this baseline is the same as the one for the FRNN-MGE~\cite{basaldella2018bidirectional}. The RNN used here is a bidirectional GRU with one hidden layer and the size of the hidden layer is $256$.
\subsubsection{Forward RNN with medical vector embedding (FRNN-MVE)} We split the sequence into subsequences with equal interval~$l$. The input at each step is the vector representation of the medical codes within the associated time interval, and the patient characteristics are appended as additional features in the final logistic regression step. Here, the RNN is a forward GRU~(or LSTM~\cite{choi2016doctor}) with one hidden layer and the size of the hidden layer is~$256$.
\subsubsection{Bidirectional RNN with medical vector embedding (BiRNN-MVE)} The inputs used for this baseline is the same as the one for the FRNN-MVE~\cite{lipton2015learning}. The RNN used here is a bidirectional GRU or LSTM~\cite{basaldella2018bidirectional} with one hidden layer and the size of the hidden layer is $256$.
\subsubsection{RETAIN} This model uses reverse time attention mechanism on RNNs for an interpretable representation of patient's EHR data~\cite{choi2016retain}. The inputs are the same as the one for FRNN-MGE, which takes the counts of medical grouping within each time interval to construct features. Similarly, the two RNNs used for generating weights are GRU-based and the size of the hidden layers are~$256$.
\subsubsection{\emph{Patient2Vec}} The inputs are the same as that for FRNN-MVE. One filter is used when generating weights for within-subsequence attention, and three filters are used for subsequence-level attention. Similarly, the RNN used here is GRU-based and there is one hidden layer and the size of the hidden layer is~$256$. 

The inputs of all baselines and \emph{Patient2Vec} are normalized to have zero mean and unit variance. We model the risk of hospitalization based on \emph{Patient2Vec} and baseline representations of patients' medical histories, and the model performance is evaluated with Area Under Curve(AUC), sensitivity, specificity, and F2-score. The validation set is used for parameter tuning and early stopping in the training process. Each experiment is repeated $20$ times and we calculate the averages and standard deviations of the above metrics, respectively.

\subsection{Experimental Results}
\label{5.5.3}
The predictive performance of \emph{Patient2Vec} and baselines are  presented in Table~\ref{tab5_result}. The results shown here for the RNN-based models are based on time interval $l=90$ days to construct subsequences.

According to Table~\ref{tab5_result}, the RNN-based models are generally capable of achieving higher prediction performance in terms of sensitivity, AUC and F2 score, except for the RNN models based on medical group embedding which have lower sensitivity. Among all RNN-based approaches, the ones based on vector embedding outperform those based on medical group embedding in terms of sensitivity, AUC, and F2 score. The bidirectional RNN models generally have higher specificity but lower sensitivity than the forward RNN models, while the bidirectional ones have comparable AUC and F2 score with the forward ones, respectively. Generally, the proposed \emph{Patient2Vec} framework outperforms the baseline methods, especially in terms of sensitivity and F2 score. 
\begin{figure*}
\centering
\includegraphics[width=0.75\textwidth]{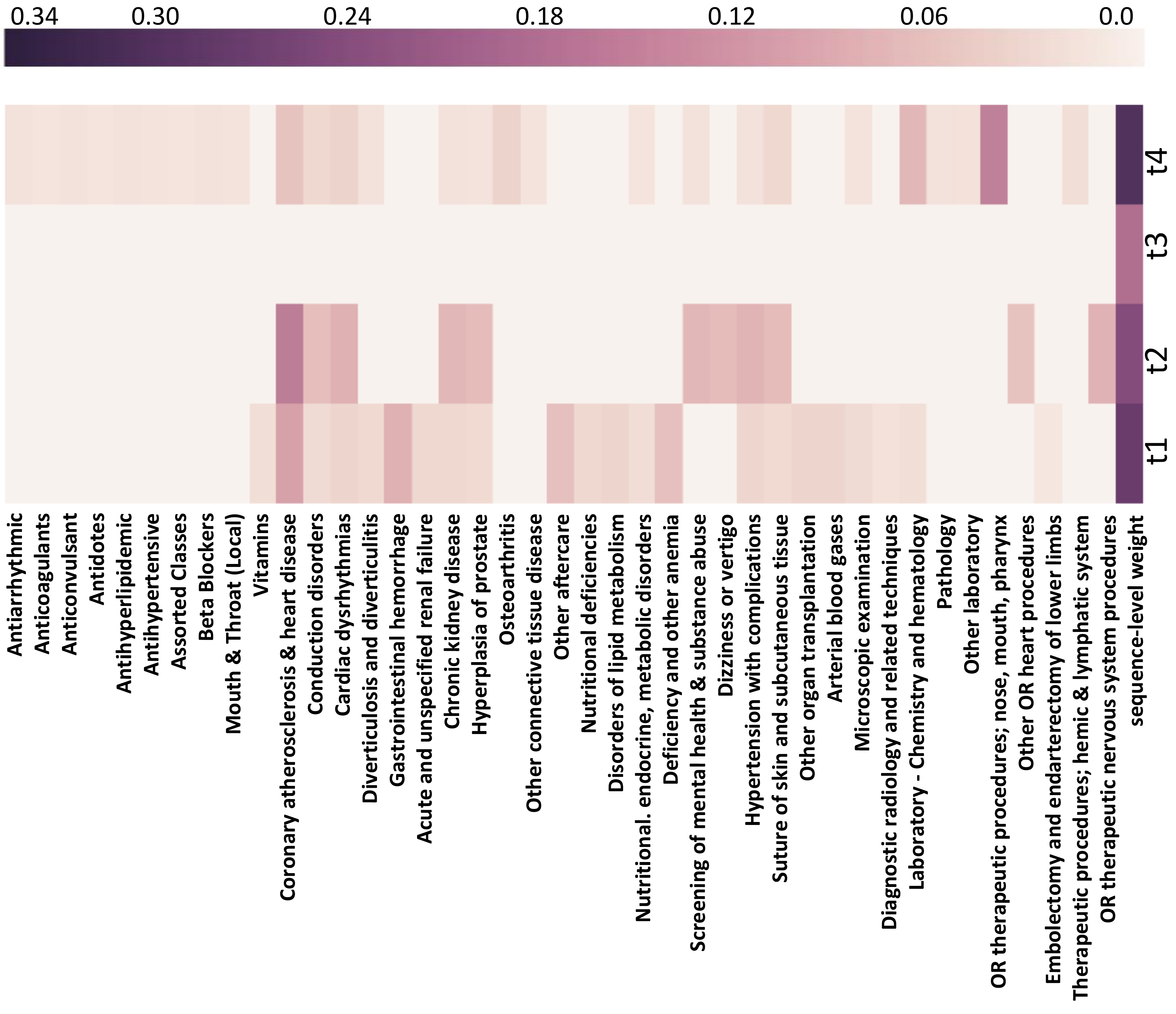}
\caption{The heat map showing feature importance for Patient A}
\label{fig5_heatmap1}
\end{figure*}

\subsection{Visualization \& Interpretation}
In addition to predictive performance, we interpret the learned representation by understanding the relative importance of clinical events in a patient's EHR data. 
Considering the feature importance learned by \emph{Patient2Vec} are personalized for an individual patient, we illustrate it with two example patients. 
Figures~\ref{fig5_pta} and \ref{fig5_ptb} present the profiles of two individuals, Patient A and Patient B, respectively. To facilitate the interpretation, instead of using raw medical codes, we present the clinical groups from the AHRQ clinical classification software on diagnoses and procedure codes, as well as pharmaceutical groups for medications.

\begin{table}[!htbp]
\centering
\caption{The top clinical groups with high weights in hospitalized patients}
\label{tab5_topfeatures}
\begin{tabular}{c p{6.2cm}}
\hline
Index & Clinical Groups \vspace{3pt}\\ \hline 
Diagnoses & \vspace{3pt}\\
1 & Essential hypertension \vspace{3pt}\\
2 & Other connective tissue disease\vspace{3pt}\\
3 & Spondylosis; intervertebral disc disorders; other back problems\vspace{3pt}\\
4 & Other lower respiratory disease\vspace{3pt}\\
5 & Disorders of lipid metabolism\vspace{3pt}\\
6 & Other aftercare\vspace{3pt}\\
7 & Diabetes mellitus without complication\vspace{3pt}\\
8 & Screening and history of mental health and substance abuse codes\vspace{3pt}\\
9 & Other nervous system disorders\vspace{3pt}\\
10 & Other screening for suspected conditions (not mental disorders or infectious disease)\vspace{3pt}\\
\hline 
Procedures & \vspace{3pt}\\ 
1 & Other OR therapeutic procedures on nose; mouth and pharynx\vspace{3pt}\\ 
2 & Suture of skin and subcutaneous tissue\vspace{3pt}\\ 
3 & Other therapeutic procedures on eyelids; conjunctiva; cornea\vspace{3pt}\\ 
4 & Laboratory - Chemistry and hematology\vspace{3pt}\\ 
5 & Other laboratory\vspace{3pt}\\ 
6 & Other OR therapeutic procedures of urinary tract\vspace{3pt}\\ 
7 & Other OR procedures on vessels other than head and neck\vspace{3pt}\\ 
8 & Therapeutic radiology for cancer treatment\vspace{3pt}\\ 
\hline
Medications & \vspace{3pt}\\
1 & Diagnostic Products \vspace{3pt}\\
2 & Analgesics-Narcotic \vspace{3pt}\\
\hline
\end{tabular}
\end{table}

\Figure[t!](topskip=0pt, botskip=0pt, midskip=0pt)[width=.48\textwidth]{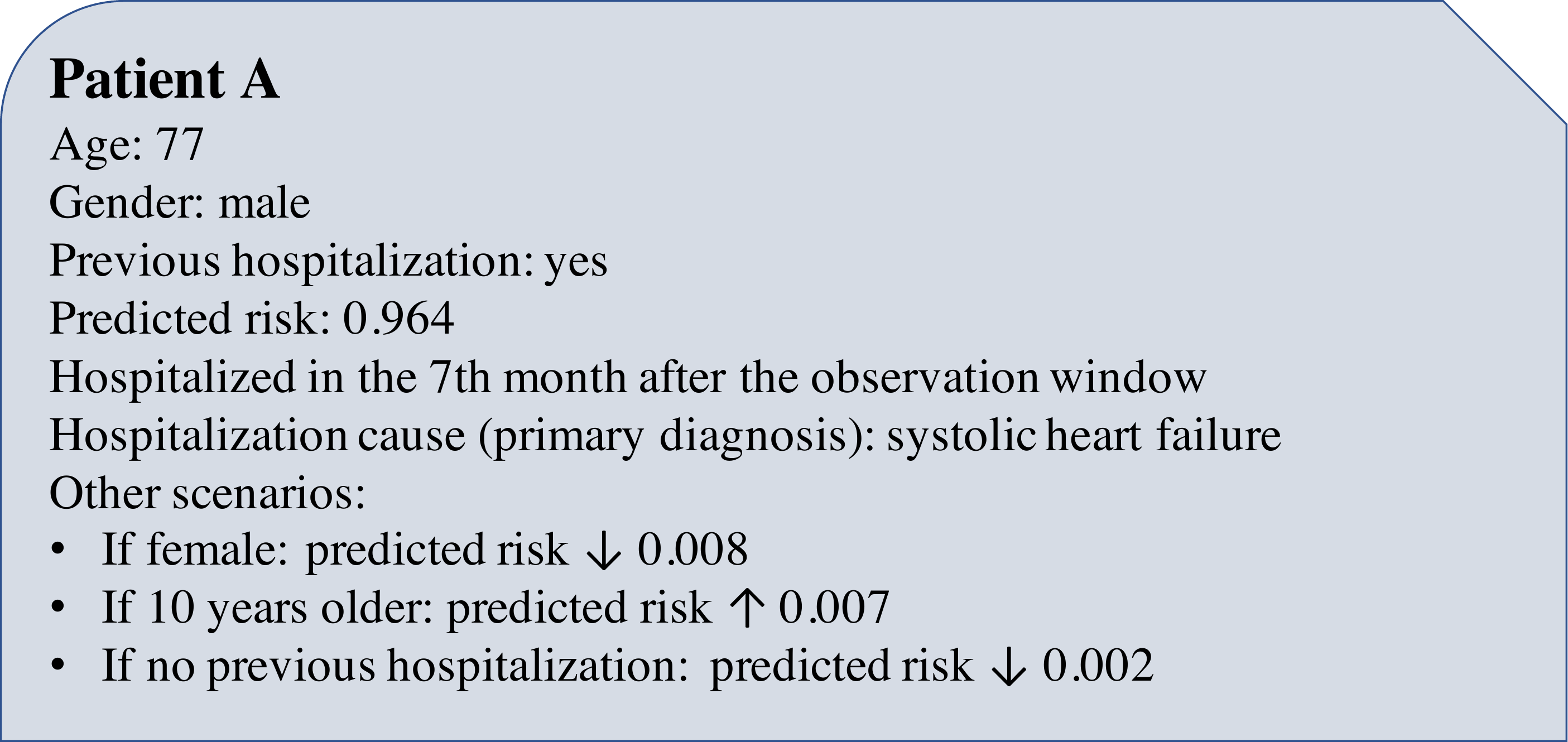}
{The profile of Patient A.\label{fig5_pta}}

\Figure[t!](topskip=0pt, botskip=0pt, midskip=0pt)[width=.48\textwidth]{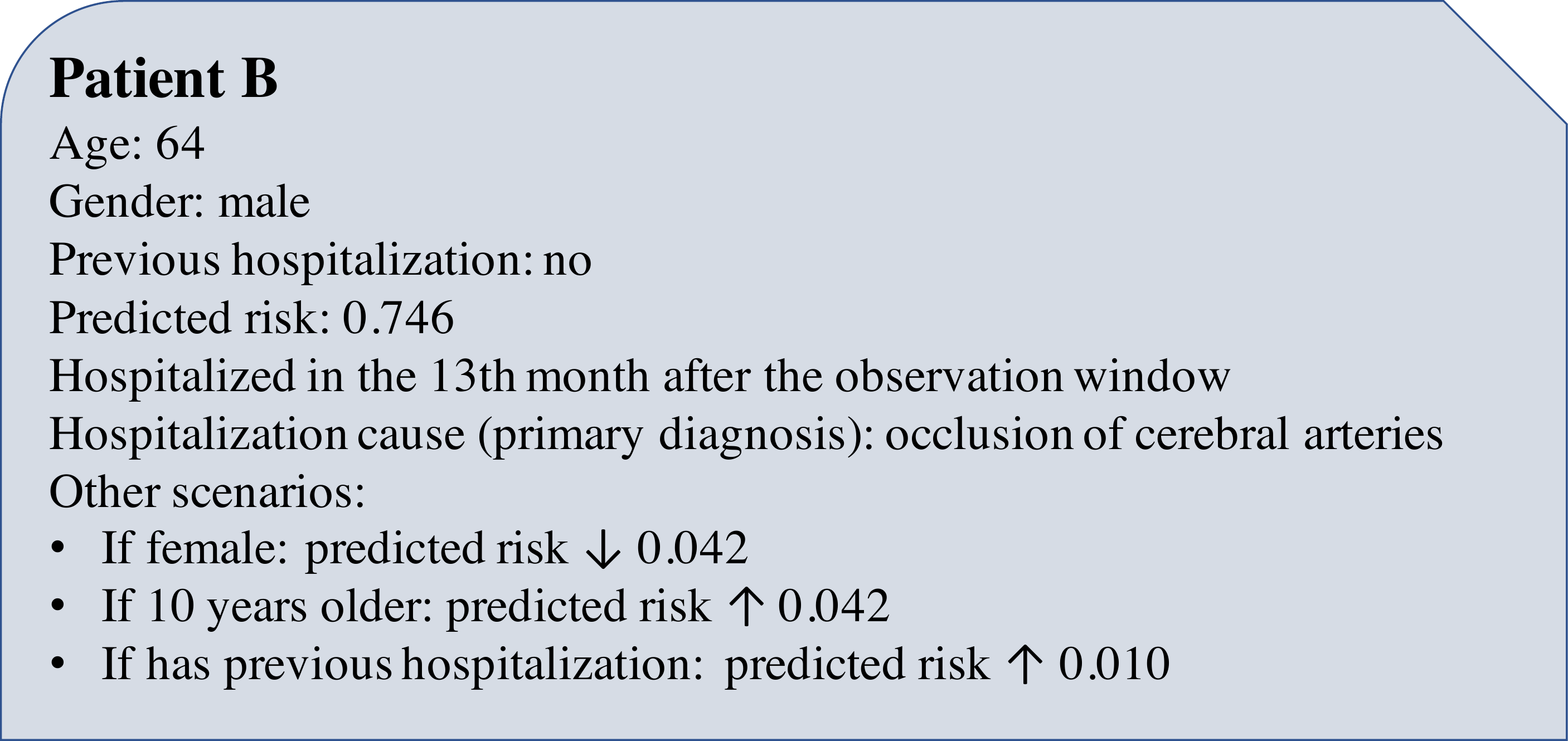}{The profile of Patient B.\label{fig5_ptb}}

According to Figure~\ref{fig5_pta}, Patient A is a male patient who has hospitalization history in the observation window and is admitted to the hospital seven months after the end of the observation window for congestive heart failure. The predicted risk is $96.4\%$, while the risk decreases for female patients or patients without hospitalization history. It is also not surprising to observe an increased risk for older patients. 
The heat map in Figure~\ref{fig5_heatmap1} shows the relative importance of the medical events in this patient's medical record at each time window and the first row of the heat map presents the subsequence-level attention. The darker color indicates a stronger correlation between the clinical events and the outcome. %%Kamran make sure that i refer to t1, t2, t3 and t4 appropriately- assuming they need to use another format
Accordingly, we observe that the last subsequence, t4, is the most important with respect to hospitalization risk, followed by $t1$, $t2$, and $t3$ in order of importance.

Among all the clinical events in the subsequence $t4$, we observe that the \textit{OR therapeutic procedures (nose, mouth, and pharynx)}, \textit{laboratory (chemistry and hematology)}, \textit{coronary atherosclerosis \& other heart disease}, \textit{cardiac dysrhythmias}, and \textit{conduction disorders} are the ones with the highest weights, while other events such as \textit{other connected tissue disease} are less important in terms of future hospitalization risk. Additionally, some medications appear to be informative as well, including \textit{beta blockers}, \textit{antihypertensives}, \textit{anticonvulsants}, \textit{anticoagulants}, etc. In the first-time window, the medical events with high weights are \textit{coronary atherosclerosis \& other heart disease}, \textit{gastrointestinal hemorrhage}, \textit{deficiency and anemia}, and \textit{other aftercare}. In the next subsequence, the most important medical events are heart diseases and related procedures such as \textit{coronary atherosclerosis \& other heart disease}, \textit{cardiac dysrhythmias}, \textit{conduction disorders}, \textit{hypertension with complications}, \textit{other OR heart procedures}, and \textit{other OR therapeutic nervous system procedures}. We also observe that the kidney disease related diagnoses and procedures appear to be important features. Throughout the observation window, the \textit{coronary atherosclerosis \& other heart disease}, \textit{cardiac dysrhythmias}, and \textit{conduction disorders} constantly show high weights with respect to hospitalization risk, and the findings are consistent with medical literature. 

\begin{figure}
\centering
\includegraphics[width=0.5\textwidth]{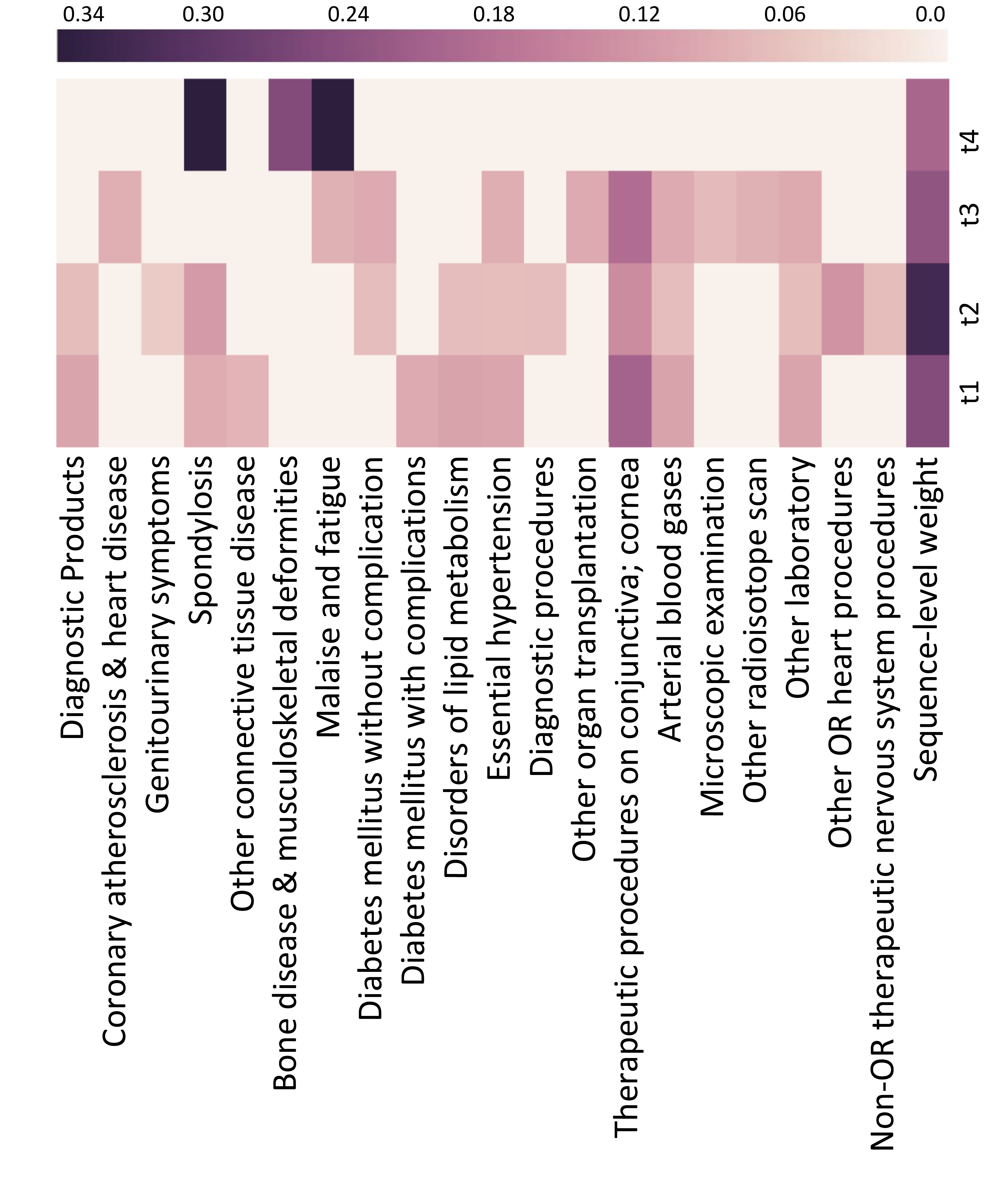}
\caption{The heat map showing feature importance for Patient B}
\label{fig5_heatmap2}
\end{figure}

Figure~\ref{fig5_ptb} presents the profile of Patient B, which is a male patient without hospitalization in the observation window. This patient is hospitalized for occlusion of cerebral arteries approximately one year after the observation window, and the predicted risk is $74.6\%$. For a similar patient who is~$10$ years older or with previous hospitalization history, the risk increases by~$4.2\%$ and~$1\%$, respectively, while there is a smaller risk of hospitalization for a female patient. To illustrate the medical events of Patient B, the heat map in Figure~\ref{fig5_heatmap2} depicts the relative importance of medical groups in the subsequences, as well as the subsequence-level weights for hospitalization risk. Similarly, the darker color indicates a stronger correlation between the clinical events and the outcome. Accordingly, we observe that the second subsequence appears to be the most important, while the last one is less predictive of future hospitalization. In fact, the medical events in the last time window are \textit{spondylosis, intervertebral disc disorders, other back problems} and \textit{other bone disease \& musculoskeletal deformities}, and \textit{malaise and fatigue}, which are not highly related to the cause of hospitalization of Patient B. 

\begin{table}[t]
\centering
\caption{The top diagnosis groups with high weights in patients hospitalized for \textit{osteoarthritis}, \textit{septicemia}, \textit{acute myocardial infarction}, \textit{congestive heart failure}, and \textit{diabetes mellitus with complications}, respectively}
\label{tab5_topfeatures_bygroups}
\begin{tabular}{c p{7cm}}
\hline
Index & Diagnosis Groups \vspace{3pt}\\ \hline 
\multicolumn{2}{l}{In patients admitted for \textit{osteoarthritis}} \vspace{3pt}\\
1 & Osteoarthritis \vspace{3pt}\\
2 & Other connective tissue disease\vspace{3pt}\\
3 & Other non-traumatic joint disorders\vspace{3pt}\\
4 & Spondylosis; intervertebral disc disorders; other back problems\vspace{3pt}\\
5 & Other aftercare\vspace{3pt}\\
\hline 
\multicolumn{2}{l}{In patients admitted for \textit{septicemia}} \vspace{3pt}\\
1 & Essential hypertension\vspace{3pt}\\ 
2 & Diabetes mellitus without complication\vspace{3pt}\\ 
3 & Disorders of lipid metabolism\vspace{3pt}\\ 
4 & Other lower respiratory disease\vspace{3pt}\\ 
5 & Other aftercare\vspace{3pt}\\
\hline
\multicolumn{2}{l}{In patients admitted for \textit{acute myocardial infarction}} \vspace{3pt}\\
1 & Coronary atherosclerosis and other heart disease \vspace{3pt}\\
2 & Medical examination/evaluation \vspace{3pt}\\ 
3 & Other screening for suspected conditions (not mental disorders or infectious disease)\vspace{3pt}\\
4 & Other lower respiratory disease\vspace{3pt}\\ 
5 & Disorders of lipid metabolism\vspace{3pt}\\ 
\hline
\multicolumn{2}{l}{In patients admitted for \textit{congestive heart failure}} \vspace{3pt}\\
1 & Congestive heart failure (nonhypertensive) \vspace{3pt}\\
2 & Coronary atherosclerosis and other heart disease \vspace{3pt}\\
3 & Cardiac dysrhythmias \vspace{3pt}\\
4 & Diabetes mellitus without complication\vspace{3pt}\\ 
5 & Other lower respiratory disease\vspace{3pt}\\ 
\hline
\multicolumn{2}{l}{In patients admitted for \textit{diabetes mellitus with complications}} \vspace{3pt}\\
1 & Diabetes mellitus with complications \vspace{3pt}\\
2 & Diabetes mellitus without complication \vspace{3pt}\\
3 & Other aftercare \vspace{3pt}\\
4 & Other nutritional; endocrine; and metabolic disorders\vspace{3pt}\\ 
5 & Fluid and electrolyte disorders\vspace{3pt}\\ 
\hline
\end{tabular}
\end{table}

In the most predictive subsequence, $t2$, we observe that \textit{other OR heart procedures}, \textit{genitourinary symptoms}, \textit{spondylosis, intervertebral disc disorders, other back problems}, \textit{therapeutic procedures on eyelid, conjunctiva, and cornea}, and \textit{arterial blood gases} have high attention weights. In the earliest time window, the most important medical events also include \textit{therapeutic procedures on eyelid, conjunctiva, and cornea}, \textit{arterial blood gases}, while \textit{diabetes}, \textit{hypertension} 
as well as diagnostic products show their relatively high importance. Throughout the observation window, medical events \textit{spondylosis, intervertebral disc disorders, other back problems}, \textit{therapeutic procedures on eyelid, conjunctiva, and cornea} are constantly with high attention weights. 
Here, diagnostic products is a medication class, which include barium sulfate, iohexol, gadopentetate dimeglumine, iodixanol, tuberculin purified protein derivative, iodixanol, regadenoson, acetone (urine), and so forth. These medications are primarily for blood or urine testing, or used as radiopaque contrast agents for x-rays or CT scans for diagnostic purposes.

Additionally, we attempt to interpret the learned representation and feature importance at the population-level. In Table~\ref{tab5_topfeatures}, we present the top $20$ clinical groups with high weights among hospitalized patients in the test set.

According to Table~\ref{tab5_topfeatures}, the most predictive diagnosis groups for future hospitalization are chronic diseases, including \textit{essential hypertension}, \textit{diabetes}, \textit{lower respiratory disease}, \textit{disorders of lipid metabolism}, and musculoskeletal diseases such as \textit{other connective tissue disease} and \textit {spondylosis, intervertebral disc disorders, other back problems}. The most important procedures are some OR therapeutic procedures and laboratory tests, such as the OR procedures on nose, mouth, and pharynx, vessels, urinary tract, eyelid, conjunctiva, cornea, etc. It is not surprising to see that diagnostic products are showing with high weights, considering these medications are used in testing or examinations for diagnostic purposes.

Moreover, we present the top diagnoses groups with high weights in patients hospitalized for different primary causes. Table~\ref{tab5_topfeatures_bygroups} shows the top $5$ diagnosis groups with high weights in patients admitted for \textit{osteoarthritis}, \textit{septicemia (except in labor)}, \textit{acute myocardial infarction}, \textit{congestive heart failure (nonhypertensive)}, and \textit{diabetes mellitus with complications}, respectively. Accordingly, we observe that the most important diagnoses for hospitalization risk prediction in population admitted for osteoarthritis are musculoskeletal diseases such as connective tissue disease, joint disorders, and spondylosis. However, the diagnoses with highest weights in the patients admitted for septicemia are chronic diseases including essential hypertension, diabetes, disorders of lipid metabolism, and respiratory disease. The top diagnoses have many overlaps between the populations admitted for acute myocardial infarction and for congestive heart failure, considering both populations are admitted for heart diseases. Here, the overlapped diagnosis groups include coronary atherosclerosis and other heart diseases and lower respiratory diseases. As for patients admitted for diabetes with complications, the top diagnoses are diabetes with or without complications, nutritional, endocrine, metabolic disorders, and fluid and electrolyte disorders. 
In general, the learned feature importance is consistent with medical literature.

\section{Discussion}
Our proposed framework is applied to the prediction of hospitalization using real EHR data that demonstrates its prediction accuracy and interpretability. This work could be further enhanced by incorporating the follow-up information on the negative patient population and investigate if it indeed shows an improved health outcome or the patient is hospitalized elsewhere.
\emph{Patient2Vec} employs a hierarchical attention mechanism, allowing us to directly interpret the weights of clinical events. In future work, we will extend the attention to incorporate demographic information for a more comprehensive and automatic interpretation. 

Although we apply \emph{Patient2Vec} to the early detection of long-term hospitalization, i.e., at least 6 months after the previous hospitalization, it could be used to predict the risk of 30-day readmission to help prevent unnecessary rehospitalizations.

\section{Conclusion}
\label{sec6}
In this paper, we propose a representation learning framework, \emph{Patient2Vec}, to learn a personalized interpretable deep representation of EHR data based on recurrent neural networks and the attention mechanism. This work improves the performance of predictive models as well as deepens the understanding of disease correlations. We apply this framework to the risk prediction of hospitalization using patients' longitudinal EHR data. The experimental results demonstrate that the proposed \emph{Patient2Vec} representation is capable of achieving a more accurate prediction than baselines approaches. Moreover, the learned feature importance in the representations are interpreted both at the individual and population levels to facilitate clinical insights.

In this work, the proposed \emph{Patient2Vec} framework is evaluated with the risk prediction of all-cause hospitalization, but in the future could be applied to predict hospitalization in more specific populations, other health related prediction problems, or domains outside of health.

 %We would like to thank anonymous reviewers for their insightful comments on the paper, as these comments led us to an improvement of the work. 



%\bibliographystyle{IEEEtran}


% \end{thebibliography}

\bibliography{refs}

\begin{thebibliography}{10}
\providecommand{\url}[1]{#1}
\csname url@samestyle\endcsname
\providecommand{\newblock}{\relax}
\providecommand{\bibinfo}[2]{#2}
\providecommand{\BIBentrySTDinterwordspacing}{\spaceskip=0pt\relax}
\providecommand{\BIBentryALTinterwordstretchfactor}{4}
\providecommand{\BIBentryALTinterwordspacing}{\spaceskip=\fontdimen2\font plus
\BIBentryALTinterwordstretchfactor\fontdimen3\font minus
  \fontdimen4\font\relax}
\providecommand{\BIBforeignlanguage}[2]{{%
\expandafter\ifx\csname l@#1\endcsname\relax
\typeout{** WARNING: IEEEtran.bst: No hyphenation pattern has been}%
\typeout{** loaded for the language `#1'. Using the pattern for}%
\typeout{** the default language instead.}%
\else
\language=\csname l@#1\endcsname
\fi
#2}}
\providecommand{\BIBdecl}{\relax}
\BIBdecl

\bibitem{yang2016hierarchical}
Z.~Yang, D.~Yang, C.~Dyer, X.~He, A.~Smola, and E.~Hovy, ``Hierarchical
  attention networks for document classification,'' in \emph{Proceedings of the
  2016 Conference of the North American Chapter of the Association for
  Computational Linguistics: Human Language Technologies}, 2016, pp.
  1480--1489.

\bibitem{kim2014convolutional}
Y.~Kim, ``Convolutional neural networks for sentence classification,''
  \emph{arXiv preprint arXiv:1408.5882}, 2014.

\bibitem{howard2018fine}
J.~Howard and S.~Ruder, ``Fine-tuned language models for text classification,''
  \emph{arXiv preprint arXiv:1801.06146}, 2018.

\bibitem{lopez2017deep}
M.~M. Lopez and J.~Kalita, ``Deep learning applied to nlp,'' \emph{arXiv
  preprint arXiv:1703.03091}, 2017.

\bibitem{cho2014learning}
K.~Cho, B.~Van~Merri{\"e}nboer, C.~Gulcehre, D.~Bahdanau, F.~Bougares,
  H.~Schwenk, and Y.~Bengio, ``Learning phrase representations using rnn
  encoder-decoder for statistical machine translation,'' \emph{arXiv preprint
  arXiv:1406.1078}, 2014.

\bibitem{nobles2018identification}
A.~L. Nobles, J.~J. Glenn, K.~Kowsari, B.~A. Teachman, and L.~E. Barnes,
  ``Identification of imminent suicide risk among young adults using text
  messages,'' in \emph{Proceedings of the 2018 CHI Conference on Human Factors
  in Computing Systems}.\hskip 1em plus 0.5em minus 0.4em\relax ACM, 2018, p.
  413.

\bibitem{kowsari2017hdltex}
K.~Kowsari, D.~E. Brown, M.~Heidarysafa, K.~J. Meimandi, M.~S. Gerber, and
  L.~E. Barnes, ``Hdltex: Hierarchical deep learning for text classification,''
  in \emph{2017 16th IEEE International Conference on Machine Learning and
  Applications (ICMLA)}, Dec 2017, pp. 364--371.

\bibitem{Kowsari2018RMDL}
K.~Kowsari, M.~Heidarysafa, D.~E. Brown, K.~Jafari~Meimandi, and L.~E. Barnes,
  ``Rmdl: Random multimodel deep learning for classification.''\hskip 1em plus
  0.5em minus 0.4em\relax ACM, 2018.

\bibitem{strobelt2018lstmvis}
H.~Strobelt, S.~Gehrmann, H.~Pfister, and A.~M. Rush, ``Lstmvis: A tool for
  visual analysis of hidden state dynamics in recurrent neural networks,''
  \emph{IEEE transactions on visualization and computer graphics}, vol.~24,
  no.~1, pp. 667--676, 2018.

\bibitem{che2018recurrent}
Z.~Che, S.~Purushotham, K.~Cho, D.~Sontag, and Y.~Liu, ``Recurrent neural
  networks for multivariate time series with missing values,'' \emph{Scientific
  reports}, vol.~8, no.~1, p. 6085, 2018.

\bibitem{luong2015effective}
M.-T. Luong, H.~Pham, and C.~D. Manning, ``Effective approaches to
  attention-based neural machine translation,'' \emph{arXiv preprint
  arXiv:1508.04025}, 2015.

\bibitem{lecun2015deep}
Y.~LeCun, Y.~Bengio, and G.~Hinton, ``Deep learning,'' \emph{Nature}, vol. 521,
  no. 7553, pp. 436--444, 2015.

\bibitem{yogatama2017generative}
D.~Yogatama, C.~Dyer, W.~Ling, and P.~Blunsom, ``Generative and discriminative
  text classification with recurrent neural networks,'' \emph{arXiv preprint
  arXiv:1703.01898}, 2017.

\bibitem{young2017recent}
T.~Young, D.~Hazarika, S.~Poria, and E.~Cambria, ``Recent trends in deep
  learning based natural language processing,'' \emph{arXiv preprint
  arXiv:1708.02709}, 2017.

\bibitem{basaldella2018bidirectional}
M.~Basaldella, E.~Antolli, G.~Serra, and C.~Tasso, ``Bidirectional lstm
  recurrent neural network for keyphrase extraction,'' in \emph{Italian
  Research Conference on Digital Libraries}, Springer.\hskip 1em plus 0.5em
  minus 0.4em\relax Springer International Publishing, 2018, pp. 180--187.

\bibitem{ghosh2016contextual}
S.~Ghosh, O.~Vinyals, B.~Strope, S.~Roy, T.~Dean, and L.~Heck, ``Contextual
  lstm (clstm) models for large scale nlp tasks,'' \emph{arXiv preprint
  arXiv:1602.06291}, 2016.

\bibitem{yue2018residual}
B.~Yue, J.~Fu, and J.~Liang, ``Residual recurrent neural networks for learning
  sequential representations,'' \emph{Information}, vol.~9, no.~3, p.~56, 2018.

\bibitem{chung2014empirical}
J.~Chung, C.~Gulcehre, K.~Cho, and Y.~Bengio, ``Empirical evaluation of gated
  recurrent neural networks on sequence modeling,'' \emph{arXiv preprint
  arXiv:1412.3555}, 2014.

\bibitem{pascanu2013difficulty}
R.~Pascanu, T.~Mikolov, and Y.~Bengio, ``On the difficulty of training
  recurrent neural networks.'' \emph{ICML (3)}, vol.~28, pp. 1310--1318, 2013.

\bibitem{wildml}
D.~Britz, ``Recurrent neural network tutorial,''
  \url{http://www.wildml.com/2015/10/}, 2015, [Accessed on October 5, 2017].

\bibitem{mnih2014recurrent}
V.~Mnih, N.~Heess, A.~Graves \emph{et~al.}, ``Recurrent models of visual
  attention,'' in \emph{Advances in neural information processing systems},
  2014, pp. 2204--2212.

\bibitem{rush2015neural}
A.~M. Rush, S.~Chopra, and J.~Weston, ``A neural attention model for
  abstractive sentence summarization,'' \emph{arXiv preprint arXiv:1509.00685},
  2015.

\bibitem{ma2018health}
T.~Ma, C.~Xiao, and F.~Wang, ``Health-atm: A deep architecture for multifaceted
  patient health record representation and risk prediction,'' in
  \emph{Proceedings of the 2018 SIAM International Conference on Data
  Mining}.\hskip 1em plus 0.5em minus 0.4em\relax SIAM, 2018, pp. 261--269.

\bibitem{rajkomar2018scalable}
A.~Rajkomar, E.~Oren, K.~Chen, A.~M. Dai, N.~Hajaj, M.~Hardt, P.~J. Liu,
  X.~Liu, J.~Marcus, M.~Sun \emph{et~al.}, ``Scalable and accurate deep
  learning with electronic health records,'' \emph{npj Digital Medicine},
  vol.~1, no.~1, p.~18, 2018.

\bibitem{lipton2015learning}
Z.~C. Lipton, D.~C. Kale, C.~Elkan, and R.~Wetzell, ``Learning to diagnose with
  lstm recurrent neural networks,'' \emph{arXiv preprint arXiv:1511.03677},
  2015.

\bibitem{che2016recurrent}
Z.~Che, S.~Purushotham, K.~Cho, D.~Sontag, and Y.~Liu, ``Recurrent neural
  networks for multivariate time series with missing values,'' \emph{arXiv
  preprint arXiv:1606.01865}, 2016.

\bibitem{johnson2016mimic}
A.~E. Johnson, T.~J. Pollard, L.~Shen, L.-w.~H. Lehman, M.~Feng, M.~Ghassemi,
  B.~Moody, P.~Szolovits, L.~A. Celi, and R.~G. Mark, ``Mimic-iii, a freely
  accessible critical care database,'' \emph{Scientific data}, vol.~3, 2016.

\bibitem{choi2016doctor}
E.~Choi, M.~T. Bahadori, A.~Schuetz, W.~F. Stewart, and J.~Sun, ``Doctor ai:
  Predicting clinical events via recurrent neural networks,'' in \emph{Machine
  Learning for Healthcare Conference}, 2016, pp. 301--318.

\bibitem{che2016interpretable}
Z.~Che, S.~Purushotham, R.~Khemani, and Y.~Liu, ``Interpretable deep models for
  icu outcome prediction,'' in \emph{AMIA Annual Symposium Proceedings}, vol.
  2016.\hskip 1em plus 0.5em minus 0.4em\relax American Medical Informatics
  Association, 2016, p. 371.

\bibitem{friedman2001greedy}
J.~H. Friedman, ``Greedy function approximation: a gradient boosting machine,''
  \emph{Annals of statistics}, pp. 1189--1232, 2001.

\bibitem{choi2016retain}
E.~Choi, M.~T. Bahadori, J.~Sun, J.~Kulas, A.~Schuetz, and W.~Stewart,
  ``Retain: An interpretable predictive model for healthcare using reverse time
  attention mechanism,'' in \emph{Advances in Neural Information Processing
  Systems}, 2016, pp. 3504--3512.

\bibitem{lin2017structured}
Z.~Lin, M.~Feng, C.~N.~d. Santos, M.~Yu, B.~Xiang, B.~Zhou, and Y.~Bengio, ``A
  structured self-attentive sentence embedding,'' \emph{arXiv preprint
  arXiv:1703.03130}, 2017.

\bibitem{cost_hcup}
C.~M. Torio and B.~J. Moore, ``National inpatient hospital costs: The most
  expensive conditions by payer, 2013,''
  \url{https://www.hcup-us.ahrq.gov/reports/statbriefs/sb204-Most-Expensive-Hospital-Conditions.jsp},
  2016.

\bibitem{wallace2014risk}
E.~Wallace, E.~Stuart, N.~Vaughan, K.~Bennett, T.~Fahey, and S.~M. Smith,
  ``Risk prediction models to predict emergency hospital admission in
  community-dwelling adults: a systematic review,'' \emph{Medical care},
  vol.~52, no.~8, p. 751, 2014.

\bibitem{de2008incidence}
E.~N. de~Vries, M.~A. Ramrattan, S.~M. Smorenburg, D.~J. Gouma, and M.~A.
  Boermeester, ``The incidence and nature of in-hospital adverse events: a
  systematic review,'' \emph{Quality and safety in health care}, vol.~17,
  no.~3, pp. 216--223, 2008.

\bibitem{purdey2012predicting}
S.~Purdey and A.~Huntley, ``Predicting and preventing avoidable hospital
  admissions: a review.'' \emph{The journal of the Royal College of Physicians
  of Edinburgh}, vol.~43, no.~4, pp. 340--344, 2012.

\bibitem{canada_admission}
H.~Ontario, ``Early identification of people at risk of hospitalization:
  Hospital admission risk prediction (harp)-a new tool for supporting providers
  and patients,'' 2013.

\bibitem{zheng_predictive_2015}
B.~Zheng, J.~Zhang, S.~W. Yoon, S.~S. Lam, M.~Khasawneh, and S.~Poranki,
  ``Predictive modeling of hospital readmissions using metaheuristics and data
  mining,'' \emph{Expert Systems with Applications}, vol.~42, no.~20, pp.
  7110--7120, 2015.

\bibitem{kansagara2011risk}
D.~Kansagara, H.~Englander, A.~Salanitro, D.~Kagen, C.~Theobald, M.~Freeman,
  and S.~Kripalani, ``Risk prediction models for hospital readmission: a
  systematic review,'' \emph{The Journal of the American Medical Association},
  vol. 306, no.~15, pp. 1688--1698, 2011.

\bibitem{giamouzis2011hospitalization}
G.~Giamouzis, A.~Kalogeropoulos, V.~Georgiopoulou, S.~Laskar, A.~L. Smith,
  S.~Dunbar, F.~Triposkiadis, and J.~Butler, ``Hospitalization epidemic in
  patients with heart failure: risk factors, risk prediction, knowledge gaps,
  and future directions,'' \emph{Journal of cardiac failure}, vol.~17, no.~1,
  pp. 54--75, 2011.

\bibitem{prescott1997gender}
E.~Prescott, A.~M. Bjerg, P.~K. Andersen, P.~Lange, and J.~Vestbo, ``Gender
  difference in smoking effects on lung function and risk of hospitalization
  for {COPD}: results from a danish longitudinal population study,''
  \emph{European Respiratory Journal}, vol.~10, no.~4, pp. 822--827, 1997.

\bibitem{ahrq}
{Agency for Healthcare Research and Quality (AHRQ)}, ``Clinical classifications
  software {(CCS)} for {ICD-9-CM},''
  \url{https://www.hcup-us.ahrq.gov/toolssoftware/ccs/ccs.jsp}, 2015.

\end{thebibliography}
\bibliographystyle{IEEEtran}
\begin{IEEEbiography}[{\includegraphics[width=1in,height=1.25in,clip,keepaspectratio]{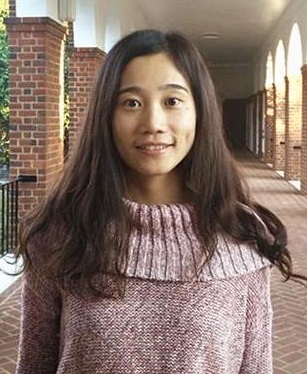}}]{\textbf{Jinghe Zhang}} ~(\href{mailto:jz4kg@virginia.edu}{jz4kg@virginia.edu}) is a lead data scientist at Target Corporation. She received her Ph.D. in Systems Engineering from the University of Virginia. Prior to entering the Ph.D. program at UVA, she received the Master of Science in Industrial and Systems Engineering from the State University of New York at Binghamton.
Her research interests are in natural language processing, machine learning, recommender systems, and health informatics. 

\end{IEEEbiography}
\begin{IEEEbiography}[{\includegraphics[width=1in,height=1.25in,clip,keepaspectratio]{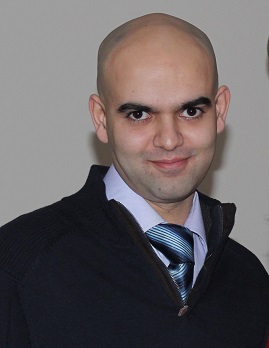}}]{\textbf{Kamran Kowsari}}~(\href{mailto:kk7nc@virginia.edu}{kk7nc@virginia.edu})  is a Ph.D. student in the Department of Systems and Information Engineering at the University of Virginia, Charlottesville,  VA. He is a member of the Sensing Systems for Health Lab. He received his Master of Science from Department of Computer Science at The George Washington  University, Washington, DC. He has more than ten years of experience in machine learning and software development. His experience includes numerous industrial and academic projects. His research interests include natural language processing, machine learning, deep learning, artificial intelligence, text mining, and unsupervised learning. 
\end{IEEEbiography}

\begin{IEEEbiography}[{\includegraphics[width=1in,height=1.25in,clip,keepaspectratio]{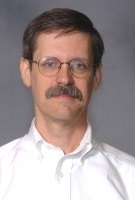}}]{\textbf{James H. Harrison, Jr.,}}(\href{mailto:jhh5y@virginia.edu}{jhh5y@virginia.\\edu}) is Associate Professor of Pathology and Director of Laboratory Information Systems at the University of Virginia Medical Center and also has appointments in the Departments of Public Health Sciences in the UVA School of Medicine, and Systems and Information Engineering in the UVA School of Engineering and Applied Sciences. He received his MD and PhD (Pharmacology) degrees from Medical University of South Carolina, Charleston, SC, completed residencies in Anatomic Pathology and Laboratory Medicine at Yale-New Haven Hospital, New Haven, CT, and completed a postdoctoral fellowship in Environmental Toxicology at Yale University, New Haven, CT. Dr. Harrison has over 25 years of experience in the field of medical informatics, including work in clinical laboratory information systems, electronic health records, clinical data analysis, and clinical data standards development.
\end{IEEEbiography}

\begin{IEEEbiography}[{\includegraphics[width=1in,height=1.25in,clip,keepaspectratio]{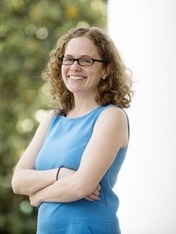}}]{\textbf{Jennifer M. Lobo}}(\href{mailto:jem4yb@virginia.edu}{jem4yb@virginia.edu}) is an  Assistant Professor of Biomedical Informatics in the Department of Public Health Sciences at University of Virginia. She received her Ph.D. in Industrial Engineering from North Carolina State University, Raleigh, NC.  Her research interests involve using mathematical modeling and stochastic optimization methods to build models that simulate the natural course of disease. These models allow for estimation of outcomes under different screening and treatment policies in the absence of randomized controlled trials, and can be used to optimize screening and treatment decisions for patients with chronic diseases.  Her projects include optimizing treatment for patients with type 2 diabetes, generating individualized decision analysis models for prostate cancer patients, and developing optimal imaging surveillance guidelines for recurrent kidney cancer.
\end{IEEEbiography}

\begin{IEEEbiography}[{\includegraphics[width=1in,height=1.25in,clip,keepaspectratio]{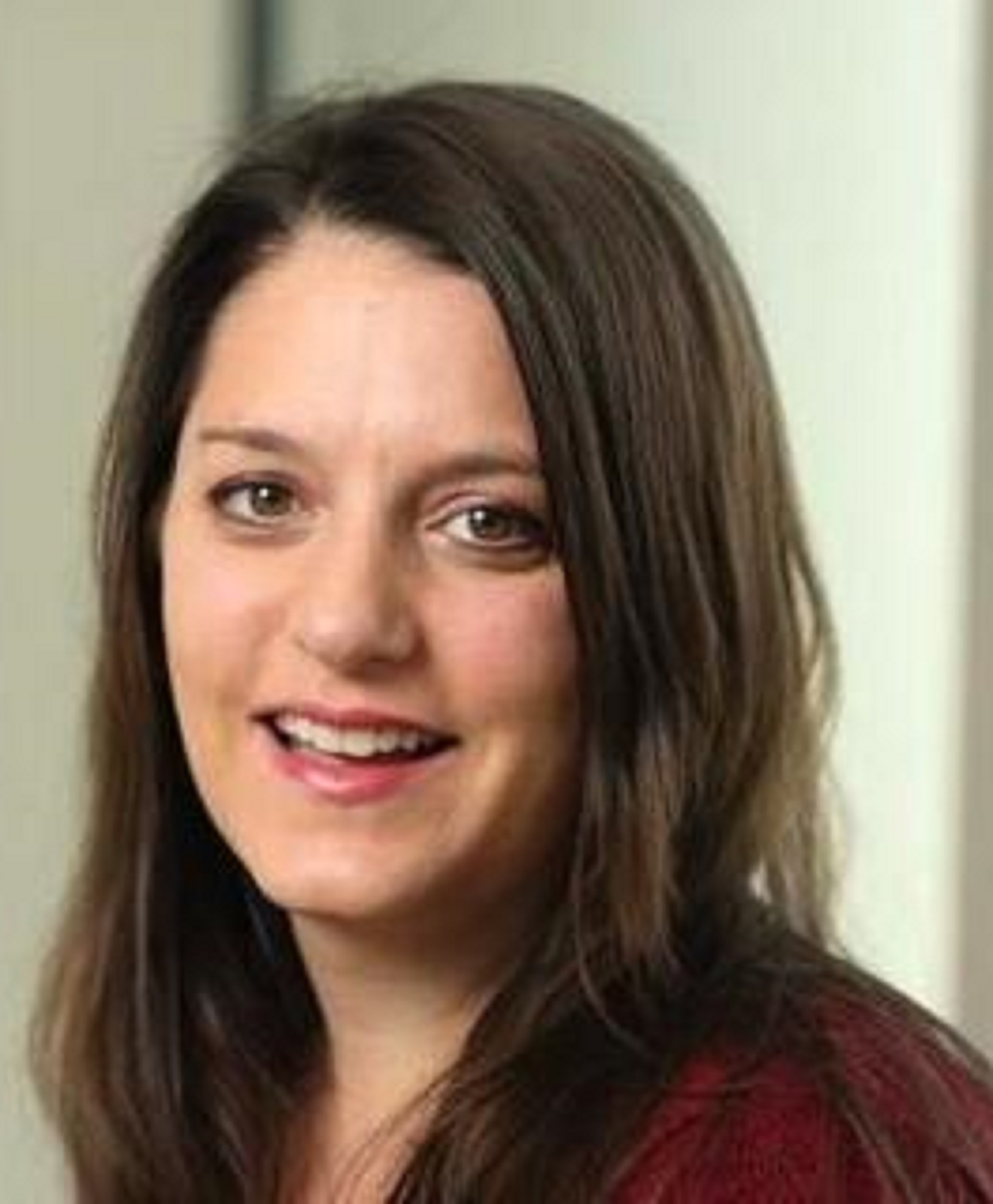}}]{\textbf{Laura E. Barnes}} (\href{mailto:lb3dp@virginia.edu}{lb3dp@virginia.edu}) is an Associate Professor in Systems and Information Engineering and the Data Science Institute at the University of Virginia. She received her Ph.D. in Computer Science from the University of South Florida, Tampa, FL. She directs the Sensing Systems for Health $(S^{2}He)$ Lab which focuses on understanding the dynamics and personalization of health and well-being through mobile sensing and analytics.
\end{IEEEbiography}

\EOD

\end{document}